\PassOptionsToPackage{table}{xcolor}
\documentclass[sigconf]{acmart}

\AtBeginDocument{%
  }

\setcopyright{none}
\renewcommand\footnotetextcopyrightpermission[1]{}

\usepackage{amsmath,amsfonts,bm}

\def\eqref#1{equation~\ref{#1}}

\def\1{\bm{1}}

\DeclareMathAlphabet{\mathsfit}{\encodingdefault}{\sfdefault}{m}{sl}
\SetMathAlphabet{\mathsfit}{bold}{\encodingdefault}{\sfdefault}{bx}{n}

\usepackage{url}
\usepackage{booktabs}
\usepackage{multirow}
\usepackage{adjustbox}
\usepackage{float}
\usepackage{placeins}
\usepackage{tabularx}
\usepackage{array}
\usepackage{subcaption}
\usepackage{enumitem}

\definecolor{ModelNameColor}{HTML}{2F5F8F}
\definecolor{OursBand}{HTML}{E7F1FB}
\definecolor{DefaultBand}{HTML}{F1F1F1}

\newcommand{\modellogo}[1]{%
  \IfFileExists{#1}{%
    \raisebox{-0.2em}{\includegraphics[height=1.25em]{#1}}%
    \hspace{0.35em}%
  }{}%
}

\newcolumntype{Y}{>{\raggedleft\arraybackslash}X}

\newcommand{\solkw}[1]{\textcolor{blue!80!black}{\textbf{#1}}}
\newcommand{\soltype}[1]{\textcolor{teal!80!black}{\textbf{#1}}}
\newcommand{\solstr}[1]{\textcolor{red!80!black}{#1}}
\newcommand{\solcomment}[1]{\textcolor{green!45!black}{#1}}
\newcommand{\solunsafe}[1]{\textcolor{red!80!black}{#1}}

\newlength{\wwwtmplen}
\NewCommandCopy{\wwworigvspace}{\vspace}
\NewDocumentCommand{\wwwnonnegvspace}{sm}{%
  \setlength{\wwwtmplen}{#2}%
  \ifdim\wwwtmplen<0pt\relax\else\IfBooleanTF{#1}{\wwworigvspace*{#2}}{\wwworigvspace{#2}}\fi}
\AtBeginEnvironment{figure*}{\let\vspace\wwwnonnegvspace}
\AtBeginEnvironment{table*}{\let\vspace\wwwnonnegvspace}

\begin{document}

\title{ContractLens: Latent Security Knowledge \\ for Malicious Smart Contract Detection}

\author{Shuyi Miao}
\affiliation{%
  \institution{Beihang University}
  \city{Beijing}
  \country{China}}

\author{Xinyi Huang}
\affiliation{%
  \institution{Beijing University of Posts and Telecommunications}
  \city{Beijing}
  \country{China}}

\author{Wangjie Qiu}
\authornote{Corresponding authors.}
\affiliation{%
  \institution{Beihang University}
  \city{Beijing}
  \country{China}}

\author{Fei Shen}
\authornotemark[1]
\affiliation{%
  \institution{National University of Singapore}
  \city{Singapore}
  \country{Singapore}}

\author{Jinchun He}
\affiliation{%
  \institution{Beihang University}
  \city{Beijing}
  \country{China}}

\author{Zhiming Zheng}
\affiliation{%
  \institution{Beihang University}
  \city{Beijing}
  \country{China}}

\author{Tat-Seng Chua}
\affiliation{%
  \institution{National University of Singapore}
  \city{Singapore}
  \country{Singapore}}

\renewcommand{\shortauthors}{Miao et al.}

\begin{abstract}

Detecting malicious smart contracts is essential to safeguarding the Web3.0 ecosystem.
However, existing auditing methods based on large language models (LLMs) largely rely on prompt engineering or attack-specific fine-tuning, while the internal representations that support malicious logic detection remain poorly characterized.
To address this gap, we propose ContractLens, a framework that localizes and uses latent security knowledge in pretrained LLMs for interpretable malicious smart contract detection.
Specifically, we first use layerwise linear probes to identify candidate layers that distinguish malicious from benign contracts, then select critical layers based on their sensitivity to malicious logic removal and stability under behaviour-preserving surface modifications.
Next, within these critical layers, we combine activation contrast, malicious-logic specificity, and probe weights into a tri-factor importance score and adaptively select malicious contract-discriminative units (MCDUs) using the knee point of the ranked score curve.
Finally, we concatenate the selected units' activations into a compact representation and train a lightweight classifier for detection, keeping the pretrained model frozen throughout.
Neuron masking experiments show that masking MCDUs causes larger drops in detection performance than masking randomly selected neurons or neurons matched in activation magnitude.
ContractLens achieves higher average detection accuracy than the evaluated methods on all three pretrained models.
Further evaluations demonstrate generalization to unseen malicious intents and show that, with the identified MCDUs fixed, a classifier trained on few labelled samples can still achieve strong detection performance. Our code is available at: \url{https://github.com/msy0513/ContractLens}.

%

\end{abstract}

\ccsdesc[500]{Security and privacy~Software and application security}
\ccsdesc[300]{Computing methodologies~Machine learning}

\keywords{Smart Contract Security, Malicious Smart Contract Detection, Large Language Models, Interpretability}

\maketitle

\section{Introduction}
\setlength{\columnsep}{5pt} 


Smart contracts~\citep{SmartContract} manage digital assets by executing
predefined rules and form core infrastructure for decentralized
finance~\citep{defi}, digital asset trading \cite{kaal2020digital}, and on-chain governance \citep{zhu2024governance}.
However, financial fraud and other malicious activities continue to
threaten this ecosystem~\citep{miao2025know,miao2026tracing,miao2026unidetect}.
Because deployed smart contracts are typically difficult to modify directly, hidden
malicious logic may continue to threaten users' assets during subsequent
interactions.
Accurately detecting such logic is therefore important for safeguarding the Web3.0 ecosystem \citep{shen2024artificial}.

\begin{figure}[t]
  \centering
   \captionsetup{skip=2pt}
  \includegraphics[width=1.0\linewidth]{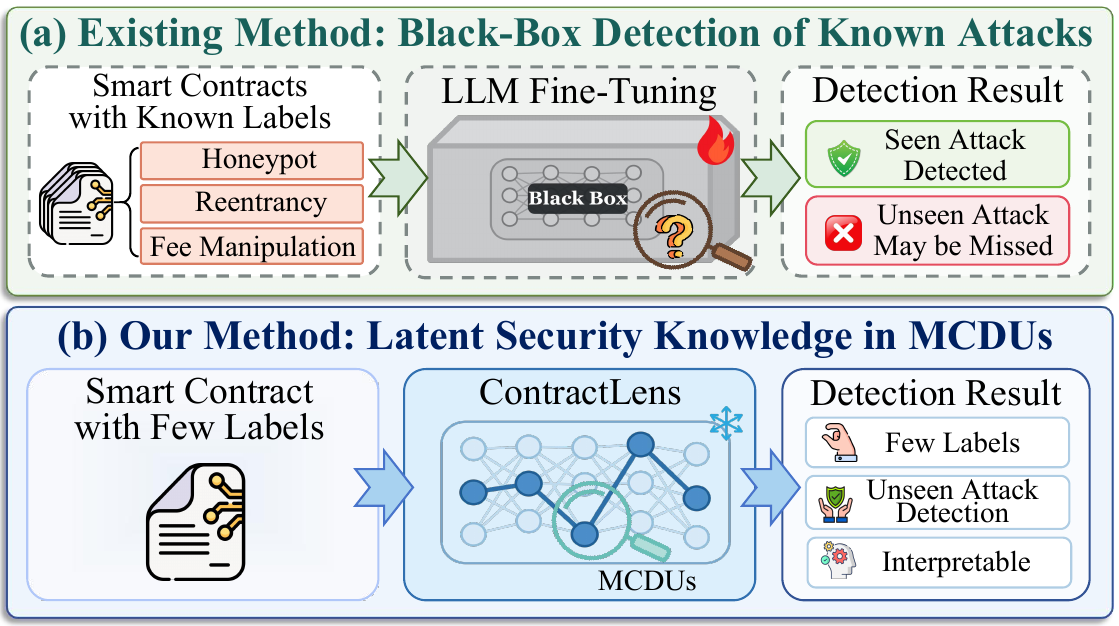}
\caption{\textbf{Latent security knowledge for interpretable smart contract auditing.}
ContractLens localizes and activates MCDUs in pretrained LLMs to detect malicious contracts.}
 \vspace{-0.6cm}
  \label{fig:motivation}
\end{figure}

Existing smart contract security analysis follows three main technical
routes: program analysis, supervised learning, and auditing with large
language models (LLMs).
Tools such as Oyente~\citep{Oyente}, Securify~\citep{Securify}, and Slither~\citep{Slither} detect violations of specified
security properties through symbolic execution, dependency analysis, or
static analysis, with their coverage
depending on the properties and rules encoded in the analyzer.
Supervised methods such as SmartIntentNN~\citep{SmartIntentNN} and its successor ~\citep{SmartIntentV2} extend
detection to malicious intent by learning mappings from contract
representations to intent labels.
More recently, LLM-based approaches ~\citep{wei2024llm,sun2024llm4vuln,ding2025smartguard,jie2025agent4vul,wei2026hkt} use prompt design, external knowledge,
and agent workflows to elicit code understanding and vulnerability
reasoning.
However, as illustrated in Figure~\ref{fig:motivation}, these learned and LLM-assisted approaches emphasize prediction quality and auditing workflows, leaving unclear which layers and neurons in pretrained models encode discriminative information and how they support malicious logic detection.


To investigate this question, we draw inspiration from how developers recognize malicious code. By studying large amounts of correct code, developers learn how programs should behave and can use this knowledge to identify malicious logic. This suggests that learning correct code can also provide a basis for malicious code detection. Similarly, LLMs exposed to extensive code during pretraining may already possess latent security knowledge relevant to distinguishing malicious from benign smart contracts. We therefore believe that such knowledge is encoded in the layers and neurons of pretrained LLMs and can be located and used for malicious smart contract detection. Our goal is to identify these internal units and use their activations for detection while keeping the pretrained model frozen.

Building on this insight, we propose ContractLens, a framework that
localizes and uses latent security knowledge in pretrained LLMs for
interpretable malicious smart contract detection.
We first train layerwise linear probes on the hidden representations of
a frozen LLM to identify candidate layers that distinguish malicious from
benign contracts.
Among these candidates, we select critical layers by testing sensitivity
to malicious logic removal and stability under behaviour-preserving
surface modifications.
Next, within the multilayer perceptron (MLP) modules of these layers, we
combine activation contrast, malicious-logic specificity, and probe weights into
a tri-factor importance score and adaptively select malicious
contract-discriminative units (MCDUs) using the knee point of the ranked
score curve.
Finally, we concatenate MCDUs activations into a compact representation
and train a lightweight classifier for detection, keeping the pretrained
model frozen throughout.
Neuron masking experiments show that masking MCDUs causes larger drops
in detection performance than masking randomly selected neurons or
neurons matched in activation magnitude.

The main contributions are summarized as follows:
\begin{itemize}
    \item We propose ContractLens, an interpretable framework that
    identifies critical layers and neurons in frozen pretrained LLMs
    and uses their latent security knowledge for malicious smart
    contract detection.

    \item We identify MCDUs through tri-factor scoring and adaptive
    selection, showing that their activations enable effective
    detection and that masking them causes larger performance drops
    than matched neuron controls.

    \item Experiments show higher average detection accuracy than
    evaluated baselines across three models, generalization to unseen
    malicious intents, and effective classifier training with few
    labelled samples using fixed MCDUs.
\end{itemize}

\section{Related Work}
\label{sec:related}

Our work connects two lines of research: LLM-based smart contract analysis and the interpretability of internal representations.
The former explores how LLMs can support malicious contract detection, while the latter examines how knowledge is encoded and used within models.
We review both directions to contextualize our study of latent security knowledge in pretrained LLMs.

\noindent\textbf{LLM-based Smart Contract Security Analysis.}
Program-analysis tools ~\citep{Oyente,Securify} detect violations of specified security properties in smart contracts.
LLM-based auditing complements these approaches by using code understanding and semantic reasoning to inspect contract behaviour.
LLM-SmartAudit~\citep{wei2024llmsmartaudit} employs specialized agents and structured analysis prompts, while LLM4Vuln~\citep{sun2024llm4vuln} systematically evaluates the effects of external vulnerability knowledge, program context, and prompt design.
SmartGuard~\citep{ding2025smartguard} combines retrieved code examples with chain-of-thought reasoning for in-context learning, and Agent4Vul~\citep{jie2025agent4vul} integrates
LLM-derived semantic information with structural program representations.
Further work explores business-guided contrastive auditing~\citep{gao2026logicscan}, knowledge distillation into lightweight auditing models~\citep{wei2026hkt}, and domain-adapted code representations for unsafe intent detection~\citep{huang2025smart}.
However, these approaches mainly focus on the generated analyses, auditing workflows, or task-specific representations, while treating the LLM as a black box. It remains underexplored which internal components of the LLM, such as layers and neurons, support malicious smart contract detection.

\noindent\textbf{Internal Representations and Neuron-level Interpretability.}
Mechanistic interpretability ~\citep{bereska2024mechanistic} investigates how information is represented and used within language models.
Linear probes~\citep{somvanshi2026bridging}reveal information decodable from intermediate representations, but predictive accuracy alone does not establish its functional relevance.
Neuron-level attribution~\citep{dai2022knowledge} localizes units associated with factual knowledge within feed-forward networks (FFNs).
Skill-neuron and sparse-probing studies~\citep{wang2022finding,gurnee2023finding} further identify small sets of task-relevant units and assess their roles through targeted interventions.
In code security, recent work~\citep{dao2026neurons} localizes FFN neurons associated with malicious Python package detection and evaluates their effects through neuron interventions.
Complementary work on explainable smart contract vulnerability detection provides evidence through expert-defined patterns~\citep{liu2021smart}, code regions~\citep{wang2026contractscanner}, and explanatory graph
substructures~\citep{yu2026contractgib}.
However, identifying pretrained LLM components encoding malicious contract logic, controlling for surface cues, and verifying their contribution to detection remain underexplored.
\section{Method}
\label{sec:method}

\begin{figure*}[!t]
    \centering
    \includegraphics[width=0.98\textwidth]{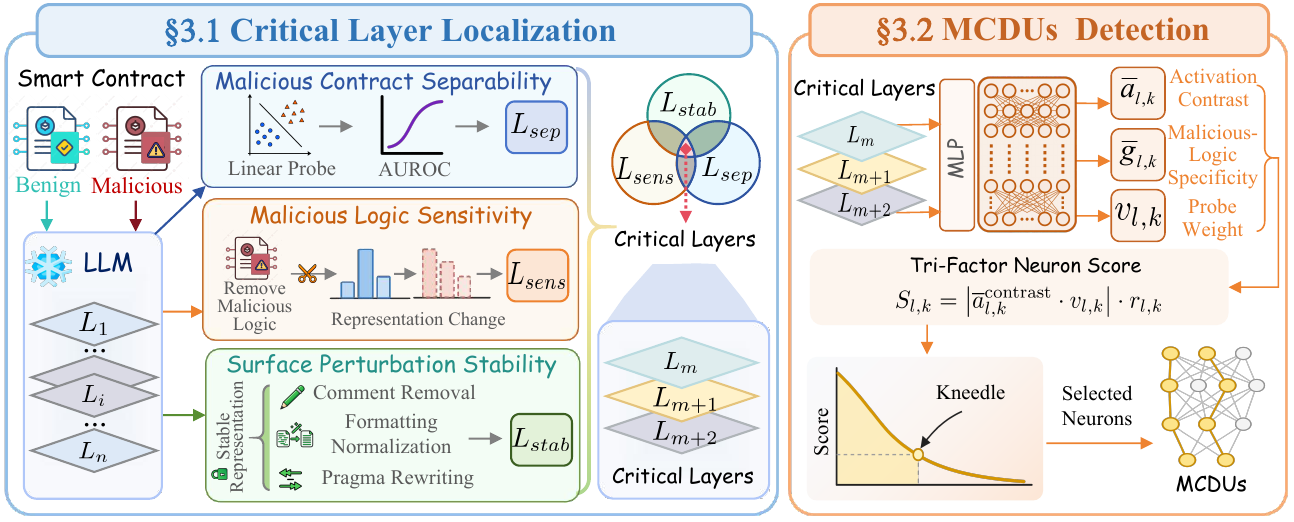}
    \vspace{-0.5cm}
    \captionsetup{skip=2pt}
    \caption{\textbf{Overview of the proposed ContractLens.}
(1) Critical layer localization: Joint evaluation of malicious smart contract separability, malicious logic sensitivity, and surface perturbation stability to locate critical layers.
(2) MCDUs detection: Tri-factor importance scoring and curvature-guided adaptive truncation within the MLP of critical layers to identify MCDUs for malicious smart contract detection using frozen pretrained LLMs.}
    \label{fig:framework}
    \vspace{-0.7cm}
\end{figure*}

As shown in Figure~\ref{fig:framework}, ContractLens localizes and extracts latent security knowledge from pretrained LLMs in two steps. 
First, it evaluates each layer for malicious contract separability, malicious logic sensitivity, and surface perturbation stability, retaining only layers that satisfy all three criteria as critical layers. 
Second, within the MLP modules of these critical layers, it identifies key neurons through tri-factor importance scoring and curvature-guided adaptive truncation. 
We refer to these neurons as MCDUs and use them for malicious smart contract detection.

\subsection{Critical Layer Localization}
\label{sec:critical-layer-localization}
For an input smart contract $x$, we obtain its layer-level representation
$\mathbf{h}_l(x)$ by mean-pooling the hidden states at each layer $l$.
We then localize critical layers using the following three criteria. 

\noindent\textbf{Malicious Contract Separability.}
We first fit a probe to the representation at each layer:
\begin{equation}
    p_l(x)
    =
    \sigma\!\left(
        \mathbf{w}_l^\top \mathbf{h}_l(x) + b_l
    \right),
    \label{eq:layer-probe}
\end{equation}
where $\sigma$ is the sigmoid function. The probe parameters
$\mathbf{w}_l$ and $b_l$ are optimized while the pretrained model remains frozen.
We evaluate each probe on the validation set and denote the AUROC score
at layer $l$ by $A^{(l)}$. Given a threshold $\tau_{\mathrm{sep}}$,
we define the separability candidate set as:
\begin{equation}
    \mathcal{L}_{\mathrm{sep}}
    =
    \left\{
        l
        \,\middle|\,
        A^{(l)} \geq \tau_{\mathrm{sep}}
    \right\}.
    \label{eq:separability-layers}
\end{equation}


High probe performance alone does not establish that a layer represents
malicious logic. A probe may instead exploit differences in variable names,
comments, formatting, contract length, or code templates. We therefore
introduce the following two criteria to determine whether the observed
separability is specifically associated with malicious behaviour.

\noindent\textbf{Malicious Logic Sensitivity.}
For each malicious contract $x$, we construct a counterfactual variant
$x_{\mathrm{cf}}$ by removing the code responsible for the malicious
behaviour while preserving the remaining code as closely as possible.
We measure the resulting representation change at layer $l$ as:
\begin{equation}
    d_{\mathrm{cf}}^{(l)}(x)
    =
    \left\lVert
        \frac{\mathbf{h}_l(x)}
             {\left\lVert\mathbf{h}_l(x)\right\rVert_2}
        -
        \frac{\mathbf{h}_l(x_{\mathrm{cf}})}
             {\left\lVert\mathbf{h}_l(x_{\mathrm{cf}})\right\rVert_2}
    \right\rVert_2.
    \label{eq:counterfactual-distance}
\end{equation}
Because removing code also changes the input structure,
$d_{\mathrm{cf}}^{(l)}$ may partly reflect the edit itself. To reduce this
effect, we compare it with representation changes caused by random
non-semantic code modifications of a similar size. The relative
sensitivity to malicious logic removal is defined as:
\begin{equation}
    R_{\mathrm{cf}}^{(l)}
    =
    \frac{
        \mathbb{E}_{x}
        \left[
            d_{\mathrm{cf}}^{(l)}(x)
        \right]
    }{
        \mathbb{E}_{x,\widetilde{x}}
        \left[
            d_{\mathrm{ctrl}}^{(l)}(x,\widetilde{x})
        \right]
        + \epsilon
    },
    \label{eq:counterfactual-ratio}
\end{equation}
where $\widetilde{x}$ is generated by randomly applying a code modification
of similar size, and $\epsilon$ prevents division by zero. We additionally
measure the decrease in the predicted malicious probability:
\[
    \Delta p^{(l)}
    =
    \mathbb{E}_{x}
    \left[
        p_l(x)-p_l(x_{\mathrm{cf}})
    \right].
\]
We define the sensitivity candidate set as:
\begin{equation}
    \mathcal{L}_{\mathrm{sens}}
    =
    \left\{
        l
        \,\middle|\,
        R_{\mathrm{cf}}^{(l)} \geq \tau_R,\;
        \Delta p^{(l)} \geq \tau_p
    \right\},
    \label{eq:sensitivity-layers}
\end{equation}
where $\tau_R$ and $\tau_p$ denote the selection thresholds for relative
representation change and malicious probability reduction, respectively.


\begin{figure*}[!t]
    \centering
    \includegraphics[width=\textwidth]{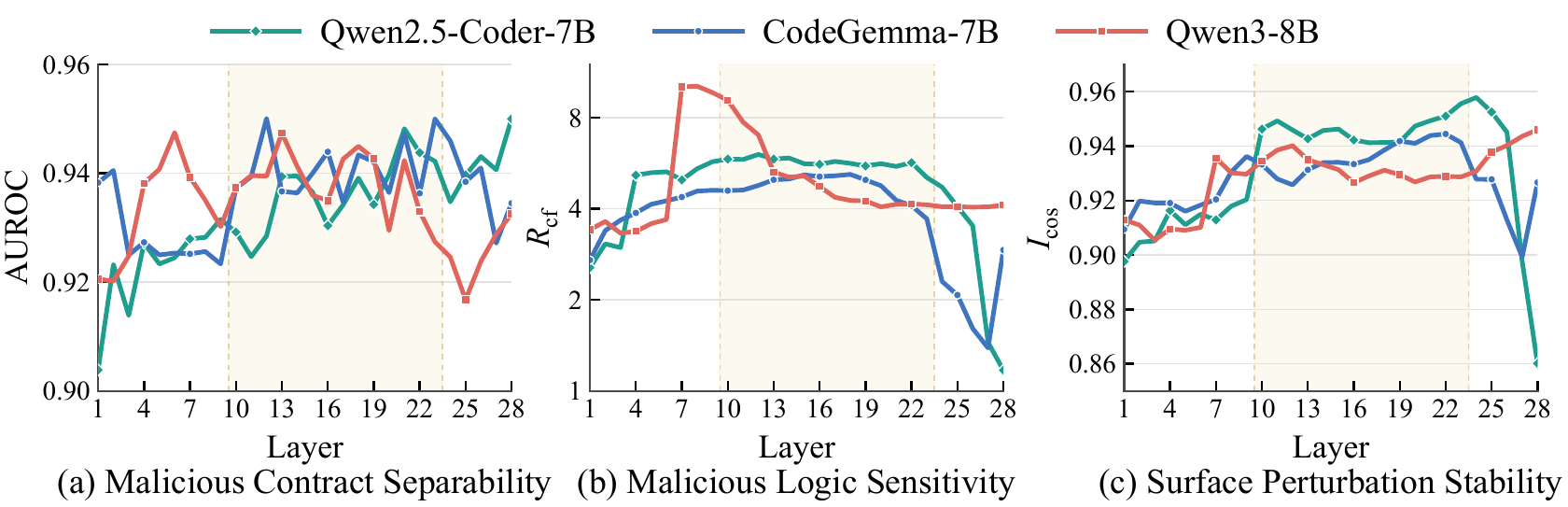}
    \vspace{-0.6cm}
    \captionsetup{skip=-2pt}
    \caption{\textbf{Visualization of layer-wise localization criteria across models.}
The malicious logic sensitivity $R_{\mathrm{cf}}^{(l)}$ measures the representation change caused by removing malicious logic.
The surface perturbation stability $I^{(l)}$ measures representation stability under code edits that preserve contract behaviour.
The linear-probe AUROC measures the ability of each layer to distinguish malicious from benign smart contracts.}
    \label{fig:threelayer}
    \vspace{-0.6cm}
\end{figure*}

\noindent\textbf{Surface Perturbation Stability.}
We apply a transformation $T$, such as comment removal, formatting
normalization, pragma rewriting, variable renaming, or function reordering,
to each smart contract $x$ without changing its program behaviour, obtaining
$x_T=T(x)$. Let $\boldsymbol{\mu}_l$ denote the mean representation of all
training contracts at layer $l$. We quantify surface perturbation stability using the following three metrics: representation similarity $I_{\mathrm{cos}}^{(l)}$, prediction
consistency $C^{(l)}$, and probability shift $S_p^{(l)}$:
\begin{equation}
\begin{aligned}
I_{\mathrm{cos}}^{(l)}
&=
\mathbb{E}\!\left[
    \cos\!\left(
        \overline{\mathbf{h}}_l(x),
        \overline{\mathbf{h}}_l(x_T)
    \right)
\right],\\
C^{(l)}
&=
\mathbb{E}\!\left[
    \mathbb{I}\!\left[
        \widehat{y}_l(x)=\widehat{y}_l(x_T)
    \right]
\right],\\
S_p^{(l)}
&=
\mathbb{E}\!\left[
    \left|p_l(x)-p_l(x_T)\right|
\right].
\end{aligned}
\label{eq:surface-stability}
\end{equation}
The stability candidate set is defined as:
\begin{equation}
    \mathcal{L}_{\mathrm{stab}}
    =
    \left\{
        l
        \,\middle|\,
        I_{\mathrm{cos}}^{(l)} \geq \tau_I,\;
        C^{(l)} \geq \tau_C,\;
        S_p^{(l)} \leq \tau_S
    \right\},
    \label{eq:stability-layers}
\end{equation}
where $\tau_I$ and $\tau_C$ are the minimum similarity and consistency,
and $\tau_S$ is the maximum probability shift.
Finally, we define the critical layers as the intersection of the three
candidate sets:
\begin{equation}
    \mathcal{L}_{\mathrm{critical}}
    =
    \mathcal{L}_{\mathrm{sep}}
    \cap
    \mathcal{L}_{\mathrm{sens}}
    \cap
    \mathcal{L}_{\mathrm{stab}}.
    \label{eq:critical-layers}
\end{equation}
As shown in Figure~\ref{fig:threelayer}, the intermediate
layers simultaneously exhibit high malicious contract separability, strong
sensitivity to malicious logic removal, and high stability under surface
modifications across all four models. 
Detailed examples of how we construct the malicious logic removal and surface-preserving variants are provided in Appendix~\ref{app:transformation_case}.
The theoretical analysis of critical layer localization and MCDUs selection is provided in Appendix~\ref{app:theoretical-analysis}.
This consistent pattern indicates that latent security knowledge for malicious smart contract detection is primarily encoded in these layers.

\subsection{Malicious Contract-Discriminative Unit Detection}
\label{sec:key-neuron-selection}

After identifying the critical layers, ContractLens further localizes the
MLP units that encode latent security knowledge. For each layer
$l\in\mathcal{L}_{\mathrm{critical}}$, we average-pool the non-padding
token activations to obtain a contract-level MLP activation vector
$\mathbf{a}_l(x)\in\mathbb{R}^{d_{\mathrm{MLP}}}$, where
$d_{\mathrm{MLP}}$ denotes the MLP width. We then fit a linear probe to
the resulting contract-level MLP activation vectors:
\begin{equation}
    q_l(x)
    =
    \sigma\!\left(
        \mathbf{v}_l^\top\mathbf{a}_l(x)+c_l
    \right),
    \label{eq:neuron-probe}
\end{equation}
where $\mathbf{v}_l\in\mathbb{R}^{d_{\mathrm{MLP}}}$ and
$c_l\in\mathbb{R}$ denote the probe weight vector and bias, respectively.

\noindent\textbf{Tri-Factor Neuron Scoring.}
Probe weights alone may not reliably identify neurons associated with
malicious logic. We therefore evaluate each neuron using three
complementary signals: activation contrast, probe weight, and malicious-logic specificity.
For neuron $k$ in layer $l$, we define the activation contrast as:
\begin{equation}
    \overline{a}_{l,k}^{\mathrm{contrast}}
    =
    \mathbb{E}_{x\sim\mathcal{D}_{\mathrm{mal}}}
    \left[a_{l,k}(x)\right]
    -
    \mathbb{E}_{x\sim\mathcal{D}_{\mathrm{ben}}}
    \left[a_{l,k}(x)\right],
    \label{eq:activation-contrast}
\end{equation}
where $\mathcal{D}_{\mathrm{mal}}$ and
$\mathcal{D}_{\mathrm{ben}}$ denote the malicious and benign training
contracts, respectively. The probe weight $v_{l,k}$ measures the
contribution of the neuron to the linear decision boundary.
To further distinguish neurons associated with malicious semantics from
those sensitive to superficial code variations, we define the
malicious-logic specificity as:
\begin{equation}
    r_{l,k}
    =
    \frac{
        \mathbb{E}_{x\sim\mathcal{D}_{\mathrm{mal}}}
        \left[
            \left|
            a_{l,k}(x)-a_{l,k}(x^{\mathrm{cf}})
            \right|
        \right]
    }{
        \mathbb{E}_{x,T}
        \left[
            \left|
            a_{l,k}(x)-a_{l,k}(x^{T})
            \right|
        \right]
        + \epsilon
    },
    \label{eq:malicious-specificity}
\end{equation}
where $\epsilon>0$ prevents division by zero. A larger $r_{l,k}$ indicates
that neuron $k$ responds more strongly to changes in malicious semantics
than to superficial code variations.
The three signals are combined into the following importance score:
\begin{equation}
    S_{l,k}
    =
    \left|
        \overline{a}_{l,k}^{\mathrm{contrast}}
        \cdot
        v_{l,k}
    \right|
    \cdot
    r_{l,k}.
    \label{eq:fusion-score}
\end{equation}
This score prioritizes neurons that respond differently to malicious and
benign contracts, contribute strongly to the probe decision, and
substantially affect the classification loss.

\noindent\textbf{MCDUs Selection.}
We collect the scores of all candidate neurons from the critical layers
and apply global min--max normalization:
\begin{equation}
    \widetilde{S}_{l,k}
    =
    \frac{
        S_{l,k}-S_{\min}
    }{
        S_{\max}-S_{\min}+\epsilon
    },
    \label{eq:score-normalization}
\end{equation}
where $S_{\min}$ and $S_{\max}$ are the minimum and maximum scores across
all candidate neurons, respectively, and $\epsilon>0$ prevents division
by zero.
We sort the normalized scores in descending order and apply
Kneedle~\citep{satopaa2011finding} to the resulting importance curve.
We retain neurons ranked before the detected knee point $K^\star$ as MCDUs. Unlike a fixed
retention threshold, this procedure determines the number of MCDUs from
the observed score distribution.

\noindent\textbf{Detection with MCDUs.}
After identifying the MCDUs, we concatenate their activations across the
critical layers and use the resulting feature vector to train a logistic
regression classifier for malicious smart contract detection. 
\section{Experiments}
\label{sec:experiments}

\begin{table*}[!t]
\centering
\vspace{-0.6cm}

\captionsetup{skip=2pt}
\caption{%
\textbf{Detection performance of malicious intent on three backbone models.}
We report AUROC (\%, $\uparrow$) and accuracy (ACC) (\%, $\uparrow$).
{Avg.} denotes the macro-average performance across ten malicious intents.
The Avg. is computed over all ten intents.
}
\label{tab:intent_comparison}

\footnotesize
\setlength{\tabcolsep}{2.0pt}
\renewcommand{\arraystretch}{1.14}

\newcommand{\metricpairplain}{%
& \multicolumn{1}{c}{AUROC}
& \multicolumn{1}{c}{ACC}%
}

\newcommand{\intentheader}{%
\multirow{2}{*}{\textbf{Method}}
& \multicolumn{2}{c}{\textbf{Avg.}}
& \multicolumn{2}{c}{\textbf{Fee}}
& \multicolumn{2}{c}{\textbf{DisableTrading}}
& \multicolumn{2}{c}{\textbf{Honeypot}}
& \multicolumn{2}{c}{\textbf{Reflection}}
& \multicolumn{2}{c}{\textbf{Blacklist}}
& \multicolumn{2}{c}{\textbf{MaxTX}}
& \multicolumn{2}{c}{\textbf{Mint}}\\
\cmidrule(lr){2-3}
\cmidrule(lr){4-5}
\cmidrule(lr){6-7}
\cmidrule(lr){8-9}
\cmidrule(lr){10-11}
\cmidrule(lr){12-13}
\cmidrule(lr){14-15}
\cmidrule(lr){16-17}
\metricpairplain
\metricpairplain
\metricpairplain
\metricpairplain
\metricpairplain
\metricpairplain
\metricpairplain
\metricpairplain\\
}

\begin{tabularx}{\textwidth}{@{}l*{16}{Y}@{}}

\midrule
\multicolumn{17}{c}{%
  \modellogo{Latex/fig/qwen-color.png}%
  \textcolor{ModelNameColor}{\textbf{Qwen2.5-Coder-7B}}%
}\\
\midrule
\intentheader
\midrule

\rowcolor{DefaultBand}
Default
& 43.68 & 50.76
& 47.84 & 57.00
& 36.64 & 48.50
& 42.43 & 47.32
& 50.24 & 56.00
& 38.96 & 50.00
& 42.12 & 53.50
& 45.76 & 48.00\\

FTSmartAudit~\citep{wei2024ftsmartaudit}
& 60.09 & 57.80
& 63.59 & 56.50
& 72.74 & 64.00
& 65.21 & 62.50
& 66.68 & 55.50
& 67.78 & 62.50
& 73.92 & 66.50
& 56.37 & 58.00\\

CodeXplain~\citep{chen2026cryptic}
& 89.46 & 81.82
& 85.57 & 80.00
& 89.02 & 81.50
& 81.92 & 79.46
& 92.00 & 84.00
& 87.29 & 79.00
& 89.10 & 79.50
& 84.91 & 74.50\\

DBESN~\citep{bedi2025dbesn}
& \underline{95.81} & \underline{91.90}
& 92.20 & 81.50
& 95.42 & \textbf{92.00}
& 94.42 & 89.29
& \textbf{97.01} & \textbf{93.50}
& 93.99 & 89.00
& \textbf{96.31} & \textbf{93.50}
& \textbf{90.71} & \textbf{85.50}\\

SCMDetector~\citep{huang2025scmdetector}
& 95.67 & 91.61
& 91.70 & 84.50
& 95.35 & 91.00
& 93.43 & \underline{91.07}
& \underline{96.84} & 89.00
& \underline{95.42} & \underline{89.50}
& \underline{96.07} & \underline{92.50}
& \underline{90.36} & 84.50\\

SmartIntentNN~\citep{SmartIntentNN}
& 93.12 & 88.61
& {93.17} & {86.27}
& {94.73} & {90.87}
& {92.48} & {90.41}
& 95.23 & \underline{93.00}
& 93.54 & 88.00
& 93.08 & 88.50
& 89.61 & {83.24}\\

SmartIntentV2~\citep{SmartIntentV2}
& 95.50 & 91.56
& \underline{93.40} & \underline{86.50}
& \underline{95.53} & \underline{91.50}
& \textbf{94.90} & \textbf{91.96}
& 96.57 & \underline{93.00}
& 94.12 & 88.00
& 93.08 & 88.50
& 90.11 & \underline{85.00}\\

\rowcolor{OursBand}
\textbf{Ours}
& \textbf{96.24} & \textbf{92.41}
& \textbf{94.85} & \textbf{87.50}
& \textbf{96.32} & \textbf{92.00}
& \underline{94.77} & \textbf{91.96}
& 96.83 & 92.00
& \textbf{96.65} & \textbf{92.00}
& 95.88 & 91.50
& 88.89 & 81.50\\

\midrule
\multicolumn{17}{c}{%
  \modellogo{Latex/fig/gemma-color.png}%
  \textcolor{ModelNameColor}{\textbf{CodeGemma-7B}}%
}\\
\midrule
\intentheader
\midrule

\rowcolor{DefaultBand}
Default
& 39.42 & 46.47
& 42.34 & 46.00
& 48.54 & 48.50
& 31.55 & 41.96
& 36.75 & 54.00
& 45.59 & 48.00
& 39.59 & 46.50
& 38.46 & 44.00\\

FTSmartAudit~\citep{wei2024ftsmartaudit}
& 66.82 & 62.14
& 56.60 & 51.50
& 70.03 & 66.00
& 60.62 & 58.93
& 59.12 & 57.00
& 62.16 & 60.50
& 70.67 & 64.00
& 64.01 & 60.50\\

CodeXplain~\citep{chen2026cryptic}
& 85.74 & 78.68
& 80.19 & 71.50
& 82.86 & 74.50
& 93.94 & 89.29
& 84.66 & 79.50
& 83.39 & 74.00
& 85.39 & 77.50
& 79.39 & 71.00\\

DBESN~\citep{bedi2025dbesn}
& \textbf{95.33} & 83.72
& \underline{92.10} & 78.00
& \textbf{95.38} & \underline{89.00}
& 94.34 & \textbf{95.54}
& \underline{95.75} & \underline{90.00}
& \underline{94.72} & 81.50
& \underline{93.83} & \underline{86.50}
& \textbf{91.67} & 79.00\\

SCMDetector~\citep{huang2025scmdetector}
& 94.42 & \underline{89.49}
& 89.20 & \textbf{85.00}
& 93.53 & 86.50
& \underline{95.18} & \underline{92.86}
& 94.20 & 87.00
& \textbf{94.78} & \textbf{88.50}
& 92.63 & \underline{86.50}
& \underline{90.98} & \underline{81.50}\\

SmartIntentNN~\citep{SmartIntentNN}
& 88.11 & 81.25
& 86.82 & 80.00
& 86.09 & 77.00
& 86.28 & 78.00
& 84.55 & 76.00
& 88.11 & 81.25
& 86.82 & 80.00
& 86.09 & 77.00\\

SmartIntentV2~\citep{SmartIntentV2}
& 86.35 & 78.01
& 78.50 & 71.00
& 86.37 & 77.50
& 87.31 & 81.25
& 83.62 & 75.00
& 82.87 & 73.50
& 84.07 & 76.50
& 82.30 & 76.00\\

\rowcolor{OursBand}
\textbf{Ours}
& \underline{95.24} & \textbf{90.94}
& \textbf{92.33} & \underline{84.00}
& \underline{94.68} & \textbf{89.50}
& \textbf{95.89} & \underline{92.86}
& \textbf{96.62} & \textbf{93.00}
& 93.98 & \underline{88.00}
& \textbf{95.14} & \textbf{91.00}
& 87.63 & \textbf{84.50}\\

\toprule
\multicolumn{17}{c}{%
  \modellogo{Latex/fig/qwen-color.png}%
  \textcolor{ModelNameColor}{\textbf{Qwen3-8B}}%
}\\
\midrule
\intentheader
\midrule

\rowcolor{DefaultBand}
Default
& 35.17 & 39.60
& 36.16 & 38.50
& 36.80 & 41.50
& 37.12 & 35.71
& 34.93 & 40.50
& 38.62 & 42.50
& 36.52 & 39.50
& 26.02 & 39.00\\

FTSmartAudit~\citep{wei2024ftsmartaudit}
& 70.45 & 67.19
& 72.07 & 67.00
& 76.87 & 66.50
& 80.84 & 75.89
& 69.93 & 64.00
& 77.49 & 67.00
& 74.30 & 70.00
& 80.51 & 78.00\\

CodeXplain~\citep{chen2026cryptic}
& 89.75 & 80.90
& 85.45 & 75.50
& 88.87 & 80.50
& 91.29 & 87.50
& 90.37 & 83.00
& 88.23 & 76.50
& 86.50 & 69.00
& 86.67 & 71.00\\

DBESN~\citep{bedi2025dbesn}
& 88.49 & 82.42
& 87.64 & 78.50
& 88.44 & 78.00
& 80.07 & 72.32
& 87.81 & 83.00
& 90.96 & 84.00
& 90.16 & 82.00
& 76.91 & 70.00\\

SCMDetector~\citep{huang2025scmdetector}
& 93.49 & 87.50
& \underline{90.99} & \underline{83.50}
& 92.84 & 84.00
& 92.38 & \underline{90.18}
& 92.93 & \underline{88.00}
& 93.39 & 85.00
& 92.93 & 83.50
& 87.56 & 77.00\\

SmartIntentNN~\citep{SmartIntentNN}
& \underline{94.25} & \underline{87.82}
& 89.86 & 83.00
& \underline{93.89} & \underline{87.00}
& \underline{94.90} & \underline{90.18}
& \underline{94.51} & 87.00
& \underline{94.42} & \underline{85.50}
& \textbf{94.10} & \underline{84.50}
& 85.90 & \underline{77.50}\\

SmartIntentV2~\citep{SmartIntentV2}
& 93.20 & 86.27
& 89.79 & \underline{83.50}
& 90.24 & 82.50
& 94.36 & 83.04
& 93.67 & 86.00
& 93.35 & 84.50
& 93.17 & 82.50
& \underline{87.67} & 76.50\\

\rowcolor{OursBand}
\textbf{Ours}
& \textbf{95.48} & \textbf{91.21}
& \textbf{92.17} & \textbf{85.00}
& \textbf{95.30} & \textbf{91.00}
& \textbf{98.57} & \textbf{91.96}
& \textbf{95.84} & \textbf{91.50}
& \textbf{94.74} & \textbf{87.00}
& \underline{94.09} & \textbf{90.50}
& \textbf{88.32} & \textbf{83.00}\\

\bottomrule
\end{tabularx}
\vspace{-0.6cm}
\end{table*}

\subsection{Implementation Details}
\noindent\textbf{Dataset.} We use the public dataset~\citep{SmartIntentNN}, which contains real-world smart contracts collected from the Binance Smart Chain (BSC) explorer. After source-code parsing and sample filtering, the dataset contains 32,141 smart contracts covering ten fine-grained intents: fee, disable trading, blacklist, reflect, max transaction, mint, honeypot, reward, rebase, and max sell. For binary detection, contracts containing at least one target intent are treated as malicious, resulting in 21,166 malicious and 10,975 benign contracts. A contract is labelled malicious based on the malicious functionality in its code, even if that functionality has not been used in an on-chain attack. Accordingly, 21,166 smart contracts are labelled as malicious, while the remaining 10,975 smart contracts are labelled as benign.
For critical-layer localization, we construct a balanced probing set by sampling 1,000 contracts annotated with at least one target intent and 1,000 benign contracts. 
%
%
For fine-grained localization, we identify MCDUs separately for each target intent by formulating ten independent one-vs-benign binary classification tasks. For each task, contracts annotated with the target intent serve as positive samples, while benign contracts serve as negative samples. To prevent data leakage, we group contracts sharing the same creator or source code and assign each group entirely to one dataset partition. For each intent, we split the resulting groups into training, validation, and test sets using a ratio of $0.64:0.16:0.20$.


\noindent\textbf{Metrics.} Following the prior work~\cite{SmartContract}, we evaluate classification performance using the area under the receiver operating characteristic curve (AUROC) and accuracy (ACC), and the $\mathrm{F}_{1}$-score.

\noindent\textbf{Baselines.}
We first assess the ability of different LLMs to detect malicious smart contracts by providing contract code directly in the prompt, and use these results as the {default} baseline.
These include two code-oriented LLMs, Qwen2.5-Coder-7B~\citep{hui2024qwen2} and CodeGemma-7B~\citep{team2024codegemma}, and one general-purpose LLM, Qwen3-8B~\citep{qwen2025qwen3}.
Then, we also compare our method with seven representative smart contract detectors.
EDSCVD~\citep{wu2024edscvd} integrates adversarial sample augmentation with dual-channel feature extraction. 
FTSmartAudit~\citep{wei2024ftsmartaudit} fine-tunes compact LLMs using knowledge distilled from stronger models for smart contract auditing. 
DBESN~\citep{bedi2025dbesn} uses a dual-branch Siamese network to capture discriminative contract features. SCMDetector~\citep{huang2025scmdetector} combines language-model-based representations with an attention-enhanced BiLSTM~\citep{siami2019performance}.
SmartIntentNN~\citep{SmartIntentNN} combines contextual sentence encoding, intent-feature highlighting, and a BiLSTM classifier. SmartIntentV2~\citep{SmartIntentV2} employs a domain-adapted pretrained model to learn smart contract representations. 
CodeXplain~\citep{chen2026cryptic} combines LLM-generated code explanations with contract code to train a supervised vulnerability detector.

\noindent\textbf{Hyperparameters.}
We set the maximum input length to 2,048 tokens and process longer contracts with sliding windows overlapping by 512 tokens. We obtain contract-level representations by mean-pooling the non-padding token embeddings. For the logistic-regression probes, we standardize the input features and search $C$ over $\{0.001, 0.01, 0.1, 1, 10\}$. 
The main comparison experiments use $C=1.0$. Critical layer selection uses $\delta_{\mathrm{AUROC}}=0.008$, $T_R=1.50$, and $T_I=0.925$.
The counterfactual probability-drop threshold is $0.02$.
The prediction-consistency threshold and maximum probability shift under surface perturbations are $0.95$ and $0.04$, respectively.
All confidence estimates use $1{,}500$ bootstrap resamples.
The detailed search ranges and selection experiments for the key
hyperparameters are provided in
Appendix~\ref{app:implementation_hyperparameters}.
All probing experiments and lightweight classifier training tasks are conducted on a single NVIDIA Tesla V100 GPU with 32\, GB of memory.
\subsection{Main Results}
\label{sec:main-results}

\noindent\textbf{Quantitative Results.} 
We evaluate ContractLens using two code-oriented LLMs: Qwen2.5-Coder-7B \citep{hui2024qwen2} and CodeGemma-7B \citep{team2024codegemma}, and one general-purpose LLM: Qwen3-8B~\citep{qwen2025qwen3}. 
Table~\ref{tab:intent_comparison} reports detection performance for the seven malicious intents with the most samples, with Avg. computed over all ten intents.
Our method achieves the highest average accuracy on all three backbone models and the highest average AUROC on Qwen2.5-Coder-7B and Qwen3-8B. 
On Qwen2.5-Coder-7B, it achieves an average AUROC of $96.24\%$ and an average accuracy of $92.41\%$. 
On CodeGemma-7B, our method obtains an average AUROC of $95.24\%$ and an average accuracy of $90.94\%$. 
On Qwen3-8B, our method achieves an average AUROC of $95.48\%$ and an average accuracy of $91.21\%$. 
Together, these results show that MCDUs encode latent security knowledge that supports detecting diverse malicious intent in smart contracts. 
Moreover, the detailed analysis of efficiency and computational cost is provided in
Appendix~\ref{app:eff}.

\noindent\textbf{Evaluation on Unseen Malicious Intents.}
To evaluate generalization to unseen malicious behaviours, we conduct a leave-one-intent-out experiment. For each held-out intent, all localization and training stages use only the remaining malicious intents and benign contracts, while the held-out intent is used exclusively for testing.
We hold out one malicious intent for testing and train the detector on the remaining intents, without fine-tuning on the held-out samples. 
\begin{figure}[t]
    \centering
     \captionsetup{skip=2pt}
    \includegraphics[
        width=\linewidth
    ]{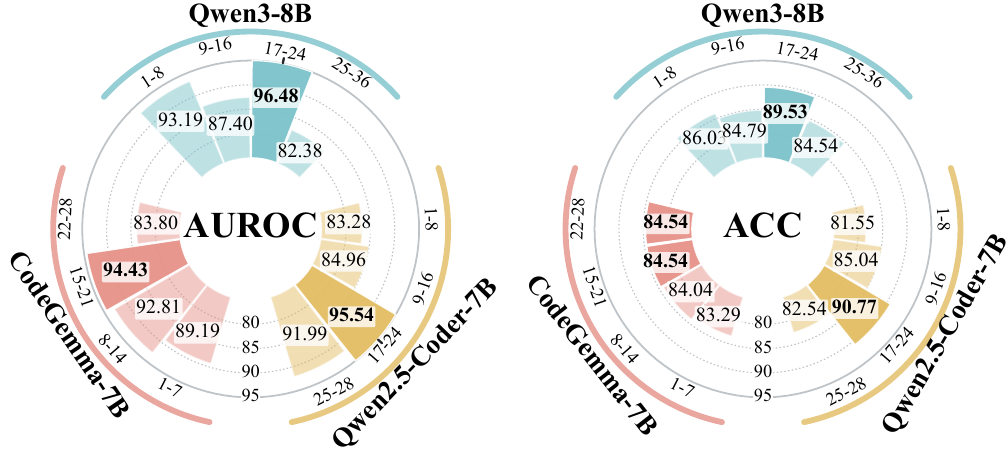}
    \caption{\textbf{Ablation study on layer depth intervals.} We partition each model into four contiguous layer intervals. The bar charts compare accuracy and AUROC across three LLMs.}
    \vspace{-0.6cm}
    \label{fig:layer_interval_ablation}
\end{figure}
Multi-label malicious smart contracts may serve as positive test samples, but each smart contract is assigned to only one held-out intent to prevent reuse across test sets. Each test set contains equal numbers of benign and malicious contracts.
Table~\ref{tab:smartdna_acc_f1} reports the results on CodeGemma-7B across five held-out intents. Our method achieves the highest accuracy and $\mathrm{F}_{1}$-score for every intent, with average scores of 73.50\% and 73.51\%, respectively. Compared with the strongest baselines, our method improves the average accuracy over SCMDetector by 7.42\% and the average $\mathrm{F}_{1}$-score over SmartIntentV2 by 10.82\%. These results show that the neuron-level discriminative signals extracted support transferable representations and robust generalization to unseen malicious intents.

\begin{table*}[!t]
\centering
\vspace{-0.6cm}
\captionsetup{skip=2pt}
\caption{%
\textbf{Detection performance on unseen malicious intents using CodeGemma-7B.}
Under the leave-one-intent-out setting, we report ACC (\%, $\uparrow$) and $\mathrm{F}_{1}$-score(\%, $\uparrow$) across five held-out malicious intents, where each intent is 
excluded from training and used only for testing in its corresponding experiment.
}
\label{tab:smartdna_acc_f1}

\footnotesize
\setlength{\tabcolsep}{1.8pt}
\renewcommand{\arraystretch}{1.08}

\begin{tabularx}{\textwidth}{@{}
>{\raggedright\arraybackslash}p{2.25cm}
*{12}{>{\centering\arraybackslash}X}
@{}}

\toprule



\multirow{2}{*}{\textbf{Method}}
& \multicolumn{2}{c}{\textbf{Fee}}
& \multicolumn{2}{c}{\textbf{MaxSell}}
& \multicolumn{2}{c}{\textbf{Reflection}}
& \multicolumn{2}{c}{\textbf{Reward}}
& \multicolumn{2}{c}{\textbf{Rebase}}
& \multicolumn{2}{c}{\textbf{Avg.}}\\

\cmidrule(lr){2-3}
\cmidrule(lr){4-5}
\cmidrule(lr){6-7}
\cmidrule(lr){8-9}
\cmidrule(lr){10-11}
\cmidrule(lr){12-13}

& \textbf{ACC} & \textbf{F1}
& \textbf{ACC} & \textbf{F1}
& \textbf{ACC} & \textbf{F1}
& \textbf{ACC} & \textbf{F1}
& \textbf{ACC} & \textbf{F1}
& \textbf{ACC} & \textbf{F1}\\

\midrule

DBESN~\citep{bedi2025dbesn}
& 58.33 & 37.50
& 53.85 & 33.33
& 60.53 & \underline{55.09}
& \underline{67.42} & 60.81
& 52.33 & 22.64
& 58.49 & 41.88\\

EDSCVD~\citep{wu2024edscvd}
& 59.38 & 36.07
& 46.15 & 36.36
& \underline{63.68} & 51.75
& 61.80 & 53.42
& 67.44 & 65.85
& 59.69 & 48.69\\

SCMDetector~\citep{huang2025scmdetector}
& 60.94 & 43.61
& \underline{73.08} & \underline{74.07}
& 51.05 & 7.92
& \underline{67.42} & 63.29
& \underline{77.91} & \underline{80.00}
& \underline{66.08} & 53.78\\

SmartIntentV2~\citep{SmartIntentV2}
& \underline{61.98} & \underline{66.36}
& 57.69 & 62.07
& 55.26 & 50.87
& 61.24 & \underline{63.87}
& 61.63 & 70.27
& 59.56 & \underline{62.69}\\

\rowcolor{OursBand}
\textbf{Ours}
& \textbf{67.19} & \textbf{70.97}
& \textbf{84.62} & \textbf{84.62}
& \textbf{65.26} & \textbf{59.76}
& \textbf{70.22} & \textbf{69.71}
& \textbf{80.23} & \textbf{82.47}
& \textbf{73.50} & \textbf{73.51}\\

\bottomrule

\end{tabularx}
\vspace{-0.6cm}
\end{table*}

\subsection{Ablation Study}
\noindent\textbf{Layers Validation.}
To test whether intent-discriminative is concentrated within specific depth ranges, we conduct layer-interval ablations across three backbones.
We partition the Transformer layers of each model into four contiguous depth intervals: shallow, transition, critical, and deep.
Then, we construct the detection features exclusively from one layer interval, while keeping the classifier architecture, training data, and hyperparameters unchanged.
As shown in Figure~\ref{fig:layer_interval_ablation}, the critical intervals produce the strongest overall performance across the three backbones.
On CodeGemma-7B, AUROC increases from 89.19\% in the shallow interval to 94.43\% in the critical interval, before decreasing to 83.80\% in the deep interval.
The metric of accuracy increases from 83.29\% to 84.54\%, although the deep interval achieves the same accuracy. On Qwen3-8B, the critical interval reaches 96.48\% AUROC and 89.53\% accuracy, compared with 93.19\% and 86.02\% in the shallow interval.
On Qwen2.5-Coder-7B, the critical interval similarly achieves the highest AUROC and accuracy, reaching 95.54\% and 90.77\%, respectively.
These consistent intermediate-layer peaks support our viewpoint that latent security knowledge associated with malicious smart contract intent is concentrated within specific depth intervals rather than uniformly distributed across the network.

\begin{figure}[t]
  \centering
  \captionsetup{skip=2pt}
  \includegraphics[width=\linewidth]{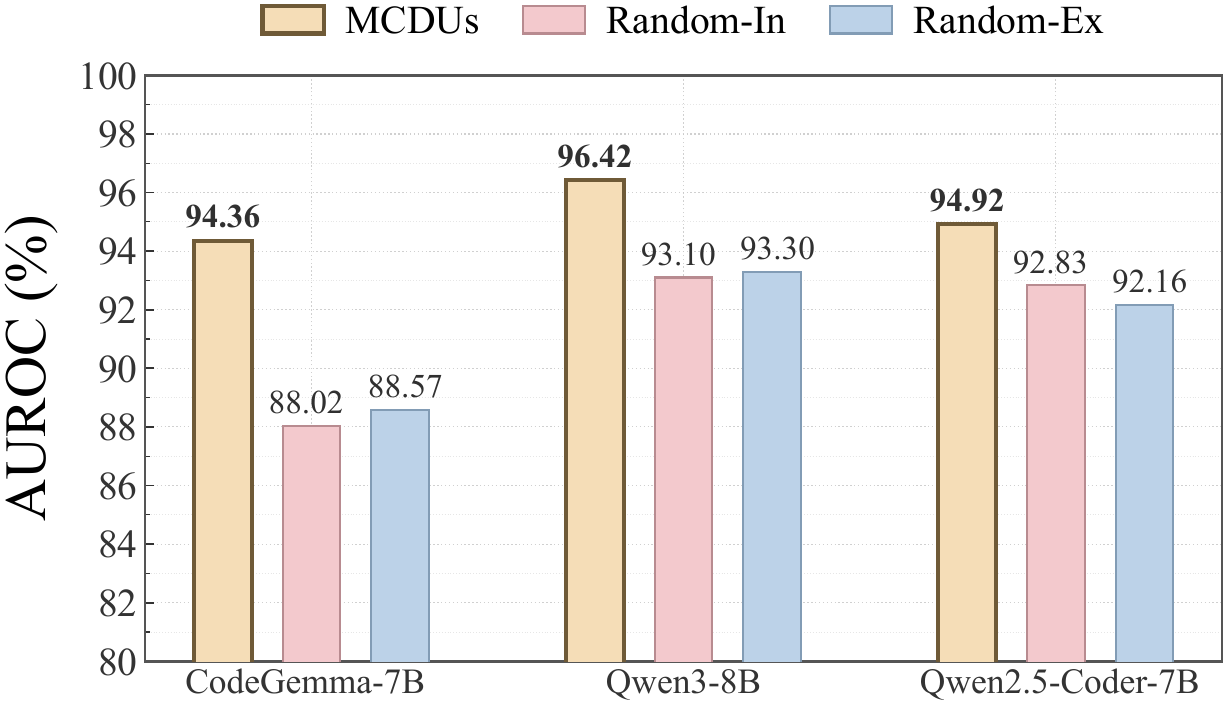}
    \caption{
    \textbf{Validation of neuron specificity.}
    We compare the AUROC of the identified MCDUs with two random selection baselines: Random-In, which randomly selects neurons from the same critical layers, and Random-Ex, which randomly selects neurons while excluding the identified MCDUs.
    }
    \vspace{-0.58cm}
  \label{fig:neuron-specificity-auroc}
\end{figure}

\noindent\textbf{Neurons Validation.} 
To determine whether detection performance depends on the identified neurons rather than arbitrary neurons within the critical layers, we compare MCDUs with two matched random baselines.
Each random baseline is evaluated over 30 independent repetitions. As shown in Figure~\ref{fig:neuron-specificity-auroc}, MCDUs consistently achieve the highest test AUROC across all three models. Specifically, MCDUs obtain $94.4\%$, $96.4\%$, and $94.9\%$ on CodeGemma-7B, Qwen3-8B, and Qwen2.5-Coder-7B, respectively. Random-In achieves $88.0\%$, $93.1\%$, and $92.8\%$, whereas Random-Ex achieves $88.6\%$, $93.3\%$, and $92.2\%$. These consistent performance gaps show that randomly selecting neurons from the critical layers cannot reproduce the effectiveness of MCDUs. Instead, the results indicate that malicious intent-related latent security knowledge is concentrated in the sparse neurons identified by our method.

\begin{table}[!t]
\centering
\footnotesize
\setlength{\tabcolsep}{5pt}
\renewcommand{\arraystretch}{1.08}
\captionsetup{skip=2pt}
\caption{\textbf{Functional ablation of MCDUs through neuron masking.} 
U0 denotes randomly selected neurons, and U1 denotes non-MCDUs with similar activation magnitudes. Both sets contain the same number
of neurons as the MCDUs.
Lower performance after masking indicates a greater contribution to our smart contract detection task.}
\label{tab:mcdu-unit-ablation}

\begin{tabularx}{\columnwidth}{X|cc}
\toprule
\textbf{Method}
& \textbf{ACC ($\uparrow$)}
& \textbf{AUROC ($\uparrow$)} \\
\midrule

\rowcolor{DefaultBand}
\textit{CodeGemma-7B}
& 85.54 & 89.26 \\

\quad + Masked U0
& 85.58 {\scriptsize\textcolor{red}{(+0.04)}}
& 89.34 {\scriptsize\textcolor{red}{(+0.08)}} \\

\quad + Masked U1
& 85.54 {\scriptsize\textcolor{red}{(0.00)}}
& 89.64 {\scriptsize\textcolor{red}{(+0.38)}} \\

\rowcolor{OursBand}
\textbf{\quad + Masked MCDUs}
& \textbf{81.05} {\scriptsize\textcolor{red}{(-4.49)}}
& \textbf{86.71} {\scriptsize\textcolor{red}{(-2.55)}} \\

\specialrule{0.45pt}{2pt}{2pt}

\rowcolor{DefaultBand}
\textit{Qwen3-8B}
& 91.52 & 95.02 \\

\quad + Masked U0
& 91.52 {\scriptsize\textcolor{red}{(0.00)}}
& 95.05 {\scriptsize\textcolor{red}{(+0.03)}} \\

\quad + Masked U1
& 91.27 {\scriptsize\textcolor{red}{(-0.25)}}
& 95.05 {\scriptsize\textcolor{red}{(+0.03)}} \\

\rowcolor{OursBand}
\textbf{\quad + Masked MCDUs}
& \textbf{83.04} {\scriptsize\textcolor{red}{(-8.48)}}
& \textbf{94.17} {\scriptsize\textcolor{red}{(-0.85)}} \\

\bottomrule
\end{tabularx}
\vspace{-0.5cm}
\end{table}

\noindent\textbf{Functional Specificity \& Robustness.}
Table~\ref{tab:mcdu-unit-ablation} examines whether the identified MCDUs make a specific functional contribution to malicious smart contract detection. 
The default setting trains a linear probe on the complete MLP activations from the critical layers while keeping the backbone frozen.
``Masked U0'' masks the same number of randomly selected neurons from the critical layers. ``Masked U1'' masks non-MCDUs with similar activation magnitudes to control for activation strength.
The MCDUs directly evaluate the contribution of the neurons identified by our method.
The experiment results show that ``Masked MCDUs'' reduce accuracy and AUROC by $4.49\%$ and $2.55\%$ on {CodeGemma-7B}, and by $8.48\%$ and $0.85\%$ on Qwen3-8B. In contrast, ``Masked U0'' causes no performance degradation on either model. Under ``Masked U1'', accuracy is unchanged on CodeGemma-7B and decreases by only $0.25\%$ on Qwen3-8B, while AUROC increases slightly on both models. These controlled comparisons show that neither the number and layer locations of the masked neurons nor their activation magnitudes explain the observed degradation. The larger performance drops caused by masking MCDUs support their functional specificity across both models.

\subsection{More Results}

\begin{figure}[t]
  \centering
  \captionsetup{skip=2pt}
  \includegraphics[width=\linewidth]{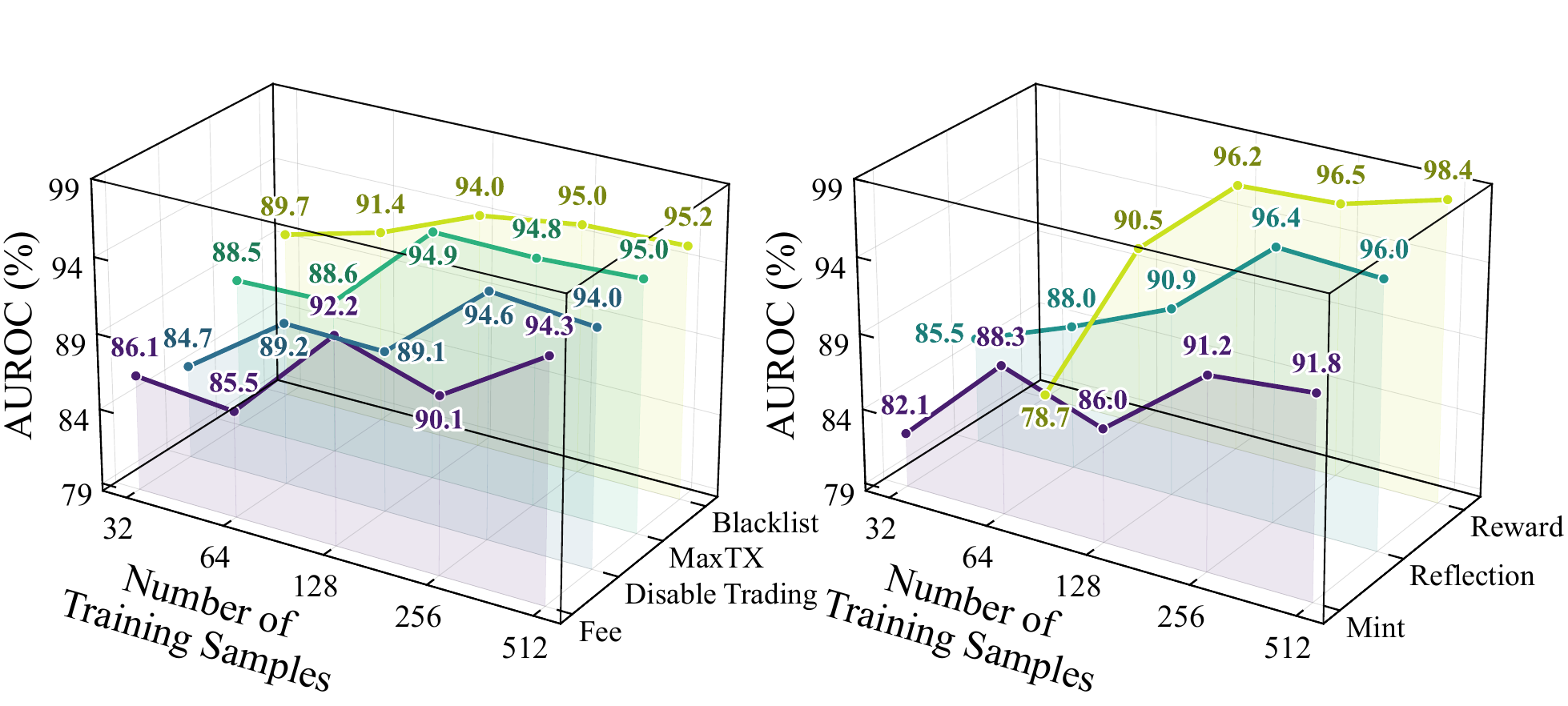}
  \caption{\textbf{Evaluation under different volumes of training data.} We report the AUROC of our method across seven malicious intents under few-shot settings.}
  \vspace{-0.6cm}
  \label{fig:traindata}
\end{figure}

\noindent\textbf{Robustness to Training-Data Size.}
Figure~\ref{fig:traindata} evaluates the data efficiency of our method under limited supervision. We freeze the CodeGemma-7B backbone, retain the same identified MCDUs, and train only a logistic-regression probe. Each training set contains equal numbers of benign and malicious contracts. With only 32 labelled contracts, our method achieves an average AUROC of $85.03\%$ across the seven malicious intents. The average AUROC increases to $88.80\%$, $91.89\%$, $94.09\%$, and $94.97\%$ as the training-set size increases, yielding an overall gain of $9.94\%$. The improvement decreases beyond 256 labelled smart contracts, indicating that performance gradually approaches saturation. The effect of additional supervision also varies across malicious intents. The AUROC increases from $78.70\%$ to $98.35\%$ for Reward and from $85.51\%$ to $96.04\%$ for Reflection. In contrast, Blacklist and MaxTX already achieve approximately $95\%$ AUROC with limited supervision and remain stable as more labelled data become available. These results show that, with the identified MCDUs fixed, limited labelled data can effectively decode their discriminative signals, supporting data-efficient classifier training without updating the backbone.

\noindent\textbf{Detection of Newly Deployed Malicious Smart Contracts.}
%
Malicious smart contracts continually evolve, requiring detectors to recognize newly deployed smart contracts and emerging attack patterns. To evaluate this capability, we conduct a temporal evaluation using a dataset collected from Dexscreener~\citep{sun2026octopusguard,dexscreener2025}. The dataset covers 14,671 BSC token addresses collected between March 18, 2023 and June 15, 2025, together with smart contract and transaction records retrieved through BscScan. After deduplication and activity filtering, it contains 1,495 representative tokens, from which the original study selects 1,201 labelled tokens, including 898 scam and 303 benign tokens. Because the collection period spans both before and after the release of Gemma-7B~\citep{gemma_2024}, we filter the dataset by smart contract deployment time. Specifically, all smart contracts used in this evaluation are deployed after the model's public release. We focus on three malicious mechanisms: honeypots, blacklist/whitelist controls, and transfer restrictions, which restrict token selling, address eligibility, and token transfers, respectively.
As shown in Table~\ref{tab:three_intent_comparison}, our method consistently outperforms the comparison methods overall, achieving the highest average accuracy of $77.50\%$ and AUROC of $84.13\%$. These results exceed the strongest competing results by $17.50\%$ in accuracy and $3.40\%$ in AUROC. Our method performs particularly well on Honeypot, reaching an accuracy of $81.25\%$ and an AUROC of $95.31\%$. It also achieves the highest accuracy of $81.25\%$ on Transfer Restriction. Overall, the results support that our method can detect newly deployed malicious smart contracts and adapt to evolving malicious behaviours without updating the backbone model.

\begin{table*}[!t]
\centering
\vspace{-0.4cm}
\captionsetup{skip=2pt}
\caption{
\textbf{Detection performance on newly deployed malicious smart contracts.}
We compare ContractLens with baseline methods on malicious smart contracts deployed after the release of Gemma-7B. We report detection accuracy and AUROC ($\%, \uparrow$) for three malicious intents, along with their averages.
}

\label{tab:three_intent_comparison}

\footnotesize
\setlength{\tabcolsep}{1.8pt}
\renewcommand{\arraystretch}{1.08}

\begin{tabularx}{\textwidth}{@{}
>{\raggedright\arraybackslash}p{2.25cm}
*{8}{>{\centering\arraybackslash}X}
@{}}

\toprule



\multirow{2}{*}{\textbf{Method}}
& \multicolumn{2}{c}{\textbf{Honeypot}}
& \multicolumn{2}{c}{\textbf{Black/Whitelist}}
& \multicolumn{2}{c}{\textbf{Transfer Restriction}}
& \multicolumn{2}{c}{\textbf{Avg.}}\\

\cmidrule(lr){2-3}
\cmidrule(lr){4-5}
\cmidrule(lr){6-7}
\cmidrule(lr){8-9}

& \textbf{ACC} & \textbf{AUROC}
& \textbf{ACC} & \textbf{AUROC}
& \textbf{ACC} & \textbf{AUROC}
& \textbf{ACC} & \textbf{AUROC}\\

\midrule

SmartIntentNN~\citep{SmartIntentNN}
& 43.75 & \underline{56.25}
& 40.00 & 32.00
& \underline{62.50} & 76.56
& 48.75 & 54.94\\

EDSCVD~\citep{wu2024edscvd}
& \underline{50.00} & 34.38
& \textbf{80.00} & \underline{72.00}
& 50.00 & 82.81
& \underline{60.00} & 63.06\\

DBESN~\citep{bedi2025dbesn}
& \underline{50.00} & 51.56
& 50.00 & \textbf{100.00}
& 50.00 & \textbf{90.63}
& 50.00 & \underline{80.73}\\

\rowcolor{OursBand}
\textbf{Ours}
& \textbf{81.25} & \textbf{95.31}
& \underline{70.00} & 68.00
& \textbf{81.25} & \underline{89.06}
& \textbf{77.50} & \textbf{84.13}\\

\bottomrule

\end{tabularx}

\vspace{-0.5cm}
\end{table*}

\section{Case Study: A Real Honeypot–Mint Token}
\label{app:realcase}

\providecommand{\solkw}[1]{\textcolor{blue!60!black}{#1}}
\providecommand{\soltype}[1]{\textcolor{purple!60!black}{#1}}
\providecommand{\solunsafe}[1]{\textcolor{red!70!black}{#1}}
\providecommand{\pos}[1]{\textcolor{green!50!black}{#1}}
\providecommand{\negd}[1]{\textcolor{red!70!black}{#1}}

\begin{table*}[t]
\centering
\vspace{-0.6cm}
\captionsetup{skip=2pt}
\caption{\textbf{Case study of the Lucky Shiba contract.}
The excerpt shows the transfer restriction and privileged balance update in a malicious BSC token.
Our method detects this sample, while the four baseline methods listed below miss it.
Scores represent each method's predicted probability of the malicious \textsc{Mint} label.
}
\label{tab:lucky_shiba_case}
\small
\renewcommand{\arraystretch}{1.12}
\begin{tabularx}{\textwidth}{@{}X@{}}
\toprule

\rowcolor{gray!15}
\textbf{Relevant smart contract logic} \\
\midrule

\begin{minipage}[!t]{\linewidth}
\footnotesize\ttfamily
\solkw{mapping}(\soltype{address} => \soltype{bool}) \solkw{public} allow;\par
\solkw{mapping}(\soltype{address} => \soltype{uint256}) sellOutNum;\par
\solkw{modifier} onlyOwner() \{ \solkw{require}(msg.sender == \soltype{address}(1530756684\ldots602)); \_; \}\par
\solunsafe{// Privileged access is restricted to a fixed address.}\par
\vspace{2pt}

\solkw{function} transferFrom(\soltype{address} \_from, \soltype{address} \_to, \soltype{uint256} \_value) \solkw{public} \{\par
\hspace*{1em}\ldots\par
\hspace*{1em}\solkw{if}(!allow[\_from]) \{ \solkw{require}(sellOutNum[\_from] == 0); \}\par
\solunsafe{// Non exempt addresses must have no prior delegated transfer.}\par
\hspace*{1em}\ldots\par
\hspace*{1em}sellOutNum[\_from]++;\par
\}\par
\solunsafe{// No reset appears in the source, so later transfers may fail.}\par
\vspace{2pt}

\solkw{function} addAllow(\soltype{address} holder, \soltype{bool} allowApprove) \solkw{external} onlyOwner \{ allow[holder] = allowApprove; \}\par
\solunsafe{// The privileged caller can change which holders are exempt.}\par
\vspace{2pt}

\solkw{function} mint(\soltype{address} miner, \soltype{uint256} \_value) \solkw{external} onlyOwner \{ balances[miner] = \_value; \}\par
\solunsafe{// Balance overwrite without a supply update or Transfer event.}\par
\end{minipage} \\

\addlinespace[3pt]
\midrule

\textbf{Why it is malicious.}
When \texttt{allow[\_from]} is false, \texttt{transferFrom} requires
\texttt{sellOutNum[\_from] == 0} and increments the counter after a successful transfer.
As the counter is not reset, later delegated transfers may fail and restrict token sales.
The owner can change the restriction through \texttt{addAllow} and overwrite an account balance through \texttt{mint} without updating \texttt{totalSupply}. \\

\addlinespace[3pt]
\midrule

\rowcolor{gray!15}
\textbf{Detection results.} The scores report each method's predicted probability for the malicious \textsc{Mint} label. \\

\textbf{Predicted malicious:} Our method (95.68\%). \\
\textbf{Predicted benign:} SmartIntentNN (33.70\%),
SmartIntentV2 (35.87\%), FTSmartAudit (56.22\%),
and CodeXplain (4.26\%). \\

\bottomrule
\end{tabularx}
\vspace{-0.8cm}
\end{table*}

Table~\ref{tab:lucky_shiba_case} presents a case study of the Lucky Shiba contract, which is labelled malicious in the BSC dataset. ContractLens correctly classifies it as malicious, whereas several baseline methods fail to detect it. The contract is malicious because its owner can prevent ordinary holders from selling their tokens while retaining the ability to arbitrarily modify account balances. As highlighted in Table~\ref{tab:lucky_shiba_case}, this is achieved through two mechanisms: a sell restriction implemented across \texttt{transferFrom} and \texttt{addAllow}, and a balance overwrite implemented in \texttt{mint}.

The sell restriction is implemented in \texttt{transferFrom}, which exchanges use to execute token sales. For any address not exempted via \texttt{allow}, the transfer succeeds only when \texttt{sellOutNum} is zero, and this counter is incremented after each transfer but never reset. Ordinary holders can therefore sell at most once, while only the owner can grant exemptions through the \texttt{onlyOwner}-protected \texttt{addAllow} function.
The balance overwrite is implemented in \texttt{mint}, which lets the owner directly set \texttt{balances[miner]} to an arbitrary value without updating \texttt{totalSupply} or emitting a \texttt{Transfer} event, so the change leaves no trace in the total supply or the event logs.

Neither mechanism is evident from function names or keywords alone, since \texttt{transferFrom}, \texttt{addAllow}, and \texttt{mint} are common in benign tokens. The malicious behaviour emerges only when the transfer condition, the never-reset counter, the owner-controlled exemption, and the missing supply and event updates are considered together. This case illustrates ContractLens's ability to reason about such cross-function logic and detect malicious contracts that several baseline methods miss.

\section{Conclusion}
This work investigates whether pretrained LLMs contain localized latent security knowledge that supports malicious smart contract detection without adapting the backbone. Our method localizes this knowledge at both the layer and neuron levels and uses the activations of the identified MCDUs as compact features for a lightweight classifier. Across three pretrained models, our method achieves higher average detection accuracy than all evaluated methods while keeping the backbone frozen. Masking MCDUs causes larger performance drops than masking random or activation-matched neurons, supporting their functional specificity. Our method also generalizes to unseen malicious intents and maintains strong detection performance with limited labelled data. Together, these results demonstrate that latent security knowledge in pretrained LLMs can be systematically localized and utilized for interpretable and data-efficient malicious smart contract detection.


\bibliographystyle{ACM-Reference-Format}
\bibliography{iclr2027_conference}

\appendix
\makeatletter
\g@addto@macro\normalsize{%
  \setlength\abovedisplayskip{4pt plus 1pt minus 1pt}%
  \setlength\belowdisplayskip{4pt plus 1pt minus 1pt}%
  \setlength\abovedisplayshortskip{2pt plus 1pt}%
  \setlength\belowdisplayshortskip{2pt plus 1pt}}
\makeatother
\normalsize
\renewcommand{\dbltopfraction}{0.9}
\renewcommand{\topfraction}{0.9}
\renewcommand{\dblfloatpagefraction}{0.85}
\renewcommand{\floatpagefraction}{0.85}
\setlength{\dbltextfloatsep}{8pt plus 2pt minus 2pt}
\setlength{\textfloatsep}{8pt plus 2pt minus 2pt}
\section*{Appendix}
The appendices provide additional theoretical and empirical details that support and extend the main paper. Appendix~\ref{app:theoretical-analysis} presents a theoretical analysis of critical-layer localization and MCDUs selection, including the semantic selectivity of the identified layers and the discriminative contribution of MCDUs. Appendix~\ref{app:implementation_hyperparameters} reports the implementation hyperparameters and evaluates their sensitivity. Appendix~\ref{app:eff} analyzes detection performance and per smart contract processing costs across different backbone models. Appendix~\ref{app:transformation_case} provides a case study illustrating the differences among the original smart contract, malicious logic removal, and surface only rewriting. 
Finally, Appendix~\ref{app:dis} further discusses the method and its scope.

\section{Theoretical Analysis}
\label{app:theoretical-analysis}

This section provides a theoretical analysis of critical layer localization
and MCDUs selection. We first analyze how the criteria for malicious logic
sensitivity and surface perturbation stability help identify layers that
respond to malicious logic rather than superficial code changes. We then
decompose the contribution of individual MLP units to the probe's class
separation and derive an upper bound on the loss change caused by masking
selected units. The assumptions and proofs for these results are provided
below.

\subsection{Semantic Selectivity of Critical-Layer Localization}
\label{app:semantic-selectivity}

Consider a smart contract $x$ as being determined by its behavioural semantics
$m$, including whether malicious logic is present, and surface factors $u$,
including identifier names, comments, formatting, and code organization. Let
the normalized representation at layer $l$ be
\begin{equation}
    \boldsymbol{\phi}_l(x)
    =
    \frac{\mathbf{h}_l(x)}{\|\mathbf{h}_l(x)\|_2}
    =
    \mathbf{s}_l(m)+\mathbf{n}_l(u)+\boldsymbol{\varepsilon}_l,
    \label{eq:semantic-nuisance-decomposition}
\end{equation}
where $\mathbf{s}_l$ and $\mathbf{n}_l$ denote the semantic and surface
components, respectively, and $\boldsymbol{\varepsilon}_l$ collects residual
interactions that are not captured by this additive approximation.

The counterfactual $x_{\mathrm{cf}}$ replaces the malicious semantics $m$ with
a benign counterpart $m_0$ while preserving $u$ as closely as possible. The
matched control $\widetilde{x}$ preserves $m$ and changes $u$ to
$\widetilde{u}$. Define the latent semantic and surface responses as
\begin{equation}
    D_{\mathrm{sem}}^{(l)}
    =\mathbb{E}\|\mathbf{s}_l(m)-\mathbf{s}_l(m_0)\|_2,
    \qquad
    D_{\mathrm{surf}}^{(l)}
    =\mathbb{E}\|\mathbf{n}_l(u)-\mathbf{n}_l(\widetilde{u})\|_2.
    \label{eq:latent-response-definitions}
\end{equation}
Let
$N_l=\mathbb{E}[d_{\mathrm{cf}}^{(l)}(x)]$ and
$B_l=\mathbb{E}[d_{\mathrm{ctrl}}^{(l)}(x,\widetilde{x})]$ denote the observed
counterfactual and matched-control changes used in
Equation~\ref{eq:counterfactual-ratio}.

\noindent\textbf{Theorem 1 (lower bound on semantic selectivity).}
Assume that the additive decomposition in
Equation~\ref{eq:semantic-nuisance-decomposition} holds locally and that the
expected residual changes are bounded by
$\eta_{\mathrm{cf}}^{(l)}$ and $\eta_{\mathrm{ctrl}}^{(l)}$ for the
counterfactual and control transformations, respectively. If layer $l$
satisfies
$R_{\mathrm{cf}}^{(l)}=N_l/(B_l+\epsilon)\geq\tau_R$, then
\begin{equation}
    D_{\mathrm{sem}}^{(l)}
    -\tau_R D_{\mathrm{surf}}^{(l)}
    \geq
    \tau_R\epsilon
    -\eta_{\mathrm{cf}}^{(l)}
    -\tau_R\eta_{\mathrm{ctrl}}^{(l)}.
    \label{eq:semantic-selectivity-bound}
\end{equation}
In particular, the semantic response is no smaller than the surface response,
$D_{\mathrm{sem}}^{(l)}\geq D_{\mathrm{surf}}^{(l)}$, whenever
\begin{equation}
    (\tau_R-1)D_{\mathrm{surf}}^{(l)}+\tau_R\epsilon
    \geq
    \eta_{\mathrm{cf}}^{(l)}
    +\tau_R\eta_{\mathrm{ctrl}}^{(l)}.
    \label{eq:semantic-dominance-condition}
\end{equation}

\noindent\textbf{Proof.}
By the triangle and reverse-triangle inequalities applied to
Equation~\ref{eq:semantic-nuisance-decomposition},
\begin{equation}
    N_l\leq
    D_{\mathrm{sem}}^{(l)}+\eta_{\mathrm{cf}}^{(l)},
    \qquad
    B_l\geq
    D_{\mathrm{surf}}^{(l)}-\eta_{\mathrm{ctrl}}^{(l)}.
    \label{eq:observed-latent-response-bounds}
\end{equation}
The selection condition $R_{\mathrm{cf}}^{(l)}\geq\tau_R$ gives
\begin{equation}
    D_{\mathrm{sem}}^{(l)}+\eta_{\mathrm{cf}}^{(l)}
    \geq N_l
    \geq \tau_R(B_l+\epsilon)
    \geq \tau_R\!\left(
        D_{\mathrm{surf}}^{(l)}
        -\eta_{\mathrm{ctrl}}^{(l)}+\epsilon
    \right).
\end{equation}
Rearranging proves Equation~\ref{eq:semantic-selectivity-bound}.
$\hfill\square$

Theorem~1 gives a quantitative meaning to malicious logic sensitivity: after
accounting for residual coupling, a selected layer has a provably larger
response to malicious-semantic removal than to a matched surface edit. The
remaining two criteria complete the layer-level argument. Membership in
$\mathcal{L}_{\mathrm{sep}}$ ensures that the signal is linearly readable,
whereas membership in $\mathcal{L}_{\mathrm{stab}}$ requires prediction
consistency and bounded probability shift under behaviour-preserving
transformations. Thus, the intersection in
Equation~\ref{eq:critical-layers} combines decodability, the semantic-response
bound in Theorem~1, and empirical surface invariance.

\subsection{Discriminative Contribution and Masking Effect of MCDUs}
\label{app:mcdu-perturbation}

We next analyze the unit-level probe in Equation~\ref{eq:neuron-probe}. Let
\begin{equation}
    z_l(x)=\mathbf{v}_l^\top\mathbf{a}_l(x)+c_l
    \quad\text{and}\quad
    q_l(x)=\sigma(z_l(x))
    \label{eq:unit-probe-logit}
\end{equation}
denote its logit and predicted malicious probability. Define the class-mean
logit gap at layer $l$ as
\begin{equation}
    \Gamma_l
    =
    \mathbb{E}_{x\sim\mathcal{D}_{\mathrm{mal}}}[z_l(x)]
    -
    \mathbb{E}_{x\sim\mathcal{D}_{\mathrm{ben}}}[z_l(x)].
    \label{eq:class-logit-gap}
\end{equation}

\noindent\textbf{Theorem 2 (unit contribution and masking bound).}
The class-mean logit gap decomposes exactly over the MLP units:
\begin{equation}
    \Gamma_l
    =
    \sum_{k=1}^{d_{\mathrm{MLP}}}
    v_{l,k}\,\overline{a}_{l,k}^{\mathrm{contrast}}.
    \label{eq:logit-gap-decomposition}
\end{equation}
Hence,
$|v_{l,k}\overline{a}_{l,k}^{\mathrm{contrast}}|$ is the magnitude of unit
$k$'s contribution to the class-mean separation induced by the linear probe.
Furthermore, let $\mathcal{F}$ be a set of units that are masked by setting
their activations to zero while keeping the probe fixed. The resulting logit
change is
\begin{equation}
    \Delta z_{l,\mathcal{F}}(x)
    =
    z_l(x)-z_l^{\setminus\mathcal{F}}(x)
    =
    \sum_{k\in\mathcal{F}}v_{l,k}a_{l,k}(x).
    \label{eq:masked-logit-change}
\end{equation}
Let
$\ell_{\mathrm{BCE}}(z,y)=\log(1+\exp(z))-yz$ be binary cross-entropy written
as a function of the logit. The masking-induced loss change satisfies
\begin{equation}
    \left|
        \ell_{\mathrm{BCE}}\!\left(z_l(x),y\right)
        -
        \ell_{\mathrm{BCE}}\!\left(
            z_l^{\setminus\mathcal{F}}(x),y
        \right)
    \right|
    \leq
    \left|\Delta z_{l,\mathcal{F}}(x)\right|
    \leq
    \sum_{k\in\mathcal{F}}|v_{l,k}a_{l,k}(x)|.
    \label{eq:masking-loss-bound}
\end{equation}

\noindent\textbf{Proof.}
Substituting Equation~\ref{eq:unit-probe-logit} into
Equation~\ref{eq:class-logit-gap} gives
\begin{equation}
    \Gamma_l
    =\sum_k v_{l,k}
    \left(
        \mathbb{E}_{\mathcal{D}_{\mathrm{mal}}}[a_{l,k}(x)]
        -
        \mathbb{E}_{\mathcal{D}_{\mathrm{ben}}}[a_{l,k}(x)]
    \right),
\end{equation}
because the bias $c_l$ cancels between the two class expectations. This is
Equation~\ref{eq:logit-gap-decomposition}. Zeroing the coordinates in
$\mathcal{F}$ removes exactly their terms from the logit, which proves
Equation~\ref{eq:masked-logit-change}. Finally,
\begin{equation}
    \frac{\partial\ell_{\mathrm{BCE}}(z,y)}{\partial z}
    =\sigma(z)-y,
    \qquad
    \left|\sigma(z)-y\right|\leq1.
\end{equation}
The mean-value theorem therefore gives the first inequality in
Equation~\ref{eq:masking-loss-bound}; the triangle inequality gives the
second. $\hfill\square$

\noindent\textbf{Interpretation of the MCDUs score.}
Let
\begin{equation}
    \Gamma_{l,k}
    =
    v_{l,k}\overline{a}_{l,k}^{\mathrm{contrast}}
    \label{eq:unit-logit-contribution}
\end{equation}
denotes the signed contribution of unit $k$ to the class-mean logit gap.
Equation~\ref{eq:logit-gap-decomposition} shows that
$|\Gamma_{l,k}|$ directly measures the discriminative contribution of the
unit under the linear probe.

However, a large discriminative contribution alone does not establish that
the unit responds specifically to malicious semantics, since it may also be
sensitive to superficial code variations. We therefore further weight this
contribution by the malicious-logic specificity
$r_{l,k}$ defined in Equation~\ref{eq:malicious-specificity}. The resulting
MCDUs score can be written as
\begin{equation}
    S_{l,k}
    =
    |\Gamma_{l,k}|\,r_{l,k}
    =
    \left|
        v_{l,k}\overline{a}_{l,k}^{\mathrm{contrast}}
    \right|
    r_{l,k}.
    \label{eq:exact-fusion-interpretation}
\end{equation}
Thus, the score favours units that both contribute strongly to class
separation and respond preferentially to malicious-logic changes rather than
behaviour-preserving surface variations.

The decomposition and masking bound justify the discriminative and
functional relevance of the selected units, while the malicious-logic
specificity term further constrains their semantic selectivity. Together,
these properties motivate the MCDUs scoring mechanism and the masking controls
in Table~\ref{tab:mcdu-unit-ablation}. The masking bound does not imply that
masking more units must monotonically reduce accuracy or AUROC, because
correlated units may compensate for or cancel one another. The Kneedle cutoff
is therefore used only to identify a data-adaptive boundary between the
high-score head and the long tail, rather than as a globally
optimal subset size.

\definecolor{ModelNameColor}{RGB}{38,84,124}
\definecolor{OursBand}{RGB}{235,242,250}

\begin{table}[!t]
\centering
\caption{
\textbf{Hyperparameter sensitivity analysis on CodeGemma-7B.}
Unit Jaccard measures the similarity between the selected neuron set and MCDUs identified by our method under the same setting. Blue cells indicate the final hyperparameter values selected for our model.
}
\vspace{-0.4cm}
\label{tab:codegemma_hparam_sensitivity}

\scriptsize
\setlength{\tabcolsep}{3pt}
\renewcommand{\arraystretch}{1.05}
\setlength{\arrayrulewidth}{0.5pt}

\begin{tabularx}{\columnwidth}{@{}l|c|>{\raggedright\arraybackslash}X|ccc@{}}
\toprule
\textbf{Parameter}
& \textbf{Value}
& $\boldsymbol{\mathcal{L}_{\mathrm{critical}}}$
& \begin{tabular}[b]{@{}c@{}}\textbf{Unit}\\\textbf{Jaccard}\end{tabular}
& \textbf{AUROC}
& \textbf{Accuracy}\\
\midrule

\multirow{4}{*}{$\delta_{\mathrm{AUROC}}$}
& 0.005
& $\{21\}$
& 0.25
& 0.95
& \textbf{0.90}\\
& \cellcolor{OursBand}\textbf{0.008}
& \cellcolor{OursBand}$\mathbf{\{16,18,19,21\}}$
& \cellcolor{OursBand}\textbf{1.00}
& \cellcolor{OursBand}\textbf{0.95}
& \cellcolor{OursBand}{0.84}\\
& 0.010
& $\{16,18,19,21\}$
& 1.00
& 0.95
& 0.84\\
& 0.015
& $\{10,14,15,16,18,$ $19,20,21,22\}$
& 0.33
& 0.93
& 0.84\\

\midrule
\multirow{4}{*}{$\tau_R$}
& 1.25
& $\{16,18,19,21\}$
& 1.00
& 0.95
& 0.84\\
& \cellcolor{OursBand}\textbf{1.50}
& \cellcolor{OursBand}$\mathbf{\{16,18,19,21\}}$
& \cellcolor{OursBand}\textbf{1.00}
& \cellcolor{OursBand}\textbf{0.95}
& \cellcolor{OursBand}\textbf{0.84}\\
& 1.75
& $\{16,18,19,21\}$
& 1.00
& 0.95
& 0.84\\
& 2.00
& $\{16,18,19,21\}$
& 1.00
& 0.95
& 0.84\\

\midrule
\multirow{3}{*}{$\tau_I$}
& 0.920
& $\{16,18,19,21\}$
& 1.00
& 0.95
& 0.84\\
& \cellcolor{OursBand}\textbf{0.925}
& \cellcolor{OursBand}$\mathbf{\{16,18,19,21\}}$
& \cellcolor{OursBand}\textbf{1.00}
& \cellcolor{OursBand}\textbf{0.95}
& \cellcolor{OursBand}\textbf{0.84}\\
& 0.930
& $\{18,19,21\}$
& 0.74
& 0.95
& 0.84\\

\midrule
\multirow{3}{*}{$N_{\mathrm{bootstrap}}$}
& 500
& $\{16,18,19,21\}$
& 1.00
& 0.95
& 0.84\\
& \cellcolor{OursBand}\textbf{1,500}
& \cellcolor{OursBand}$\mathbf{\{16,18,19,21\}}$
& \cellcolor{OursBand}\textbf{1.00}
& \cellcolor{OursBand}\textbf{0.95}
& \cellcolor{OursBand}\textbf{0.84}\\
& 3,000
& $\{16,18,19,21\}$
& 1.00
& 0.95
& 0.84\\

\bottomrule
\end{tabularx}
\vspace{-0.4cm}
\end{table}

\section{Implementation Hyperparameters}
\label{app:implementation_hyperparameters}

Table~\ref{tab:codegemma_hparam_sensitivity} reports a sensitivity analysis of four key hyperparameters: $\delta_{\mathrm{AUROC}}$, $\tau_R$, $\tau_I$, and $N_{\mathrm{bootstrap}}$. We use the Jaccard index to measure the similarity between the neuron set $U_{\theta}$ selected under a given configuration and the set $U_{\mathrm{default}}$ selected under the default configuration:
\begin{equation}
J(U_{\theta},U_{\mathrm{default}})
=
\frac{|U_{\theta}\cap U_{\mathrm{default}}|}
{|U_{\theta}\cup U_{\mathrm{default}}|}.
\end{equation}
The value of $J(U_{\theta},U_{\mathrm{default}})=1$ indicates that the selected neuron set is identical to the MCDUs. Across most evaluated settings, AUROC and accuracy vary only modestly, while the selected neurons remain unchanged for different values of $\tau_R$ and $N_{\mathrm{bootstrap}}$. Only larger changes in $\delta_{\mathrm{AUROC}}$ noticeably affect critical-layer and neuron selection. These results indicate that the localization and detection performance is generally robust to moderate hyperparameter variations.
Based on the validation results, we select $\delta_{\mathrm{AUROC}}=0.008$, $\tau_R=1.50$, $\tau_I=0.925$, and $N_{\mathrm{bootstrap}}=1{,}500$ as the optimal parameter configuration, which provides the best overall balance between detection performance and localization stability.
We set the thresholds for probability drop, prediction consistency, and maximum probability shift to $0.02$, $0.95$, and $0.04$, respectively. We then tune the regularization parameter $C$ of the logistic regression probe on the validation set, selecting $C=1.0$ from $\{0.001, 0.01, 0.1, 1, 10\}$.

\section{Efficiency and Computational Cost Analysis}
\label{app:eff}

\begin{figure*}[!t]
    \centering
    \includegraphics[width=\textwidth]{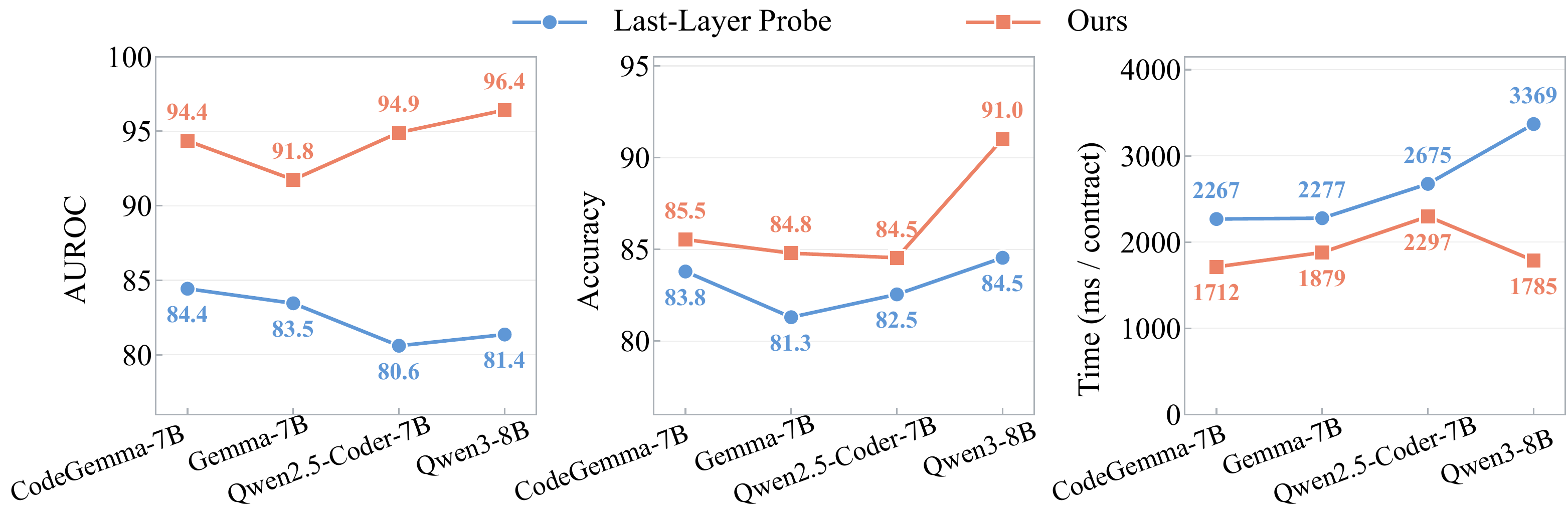}
\vspace{-2cm}
\caption{
\textbf{Detection performance and processing efficiency across backbone models.}
We compare ContractLens with the conventional last-layer probe in terms of AUROC, accuracy, and per-contract processing time. Higher AUROC and accuracy indicate better detection performance, while lower processing time indicates greater efficiency.
}
\label{fig:efficiency_analysis}
\vspace{-0.6cm}
\end{figure*}

To evaluate practical efficiency, we compare our method with the conventional last-layer probe across four backbone models under the same hardware and evaluation data. We report AUROC, accuracy, and per-smart contract processing time.
As shown in Figure~\ref{fig:efficiency_analysis}, our method consistently improves detection performance across all four backbones. Its AUROC ranges from approximately $91\%$ to $96\%$, compared with $80\%$ to $85\%$ for the last-layer probe. Our method also achieves higher accuracy on every backbone, with the largest improvement observed on Qwen3-8B. These results indicate that MCDUs provide more discriminative representations than conventional last-layer features.
Our method also requires less processing time across all evaluated backbones, with the largest reduction observed on Qwen3-8B. Thus, the performance gains do not introduce additional inference overhead. Instead, the compact representation formed by the selected MCDUs improves both detection effectiveness and processing efficiency. These results support the practical use of our method for large-scale malicious smart contract screening.

\providecommand{\solkw}[1]{\textcolor{blue!80!black}{\textbf{#1}}}
\providecommand{\soltype}[1]{\textcolor{teal!80!black}{\textbf{#1}}}
\providecommand{\solstr}[1]{\textcolor{red!80!black}{#1}}
\providecommand{\solcomment}[1]{\textcolor{green!45!black}{#1}}
\providecommand{\solunsafe}[1]{\textcolor{red!80!black}{#1}}
\section{Malicious Logic Removal and Surface Perturbation}
\label{app:transformation_case}

Table~\ref{tab:transformation_case} illustrates the two controlled code
transformations used for critical layer localization. The original input
$x$ contains an owner-controlled transfer restriction associated with the
honeypot intent. Malicious logic removal produces the counterfactual input
$x_{\mathrm{cf}}$ by removing this restriction, changing the smart contract
from malicious to benign. In contrast, a behaviour-preserving surface
perturbation produces $x_T=T(x)$ by renaming identifiers without changing
the underlying unsafe logic or its label.
These transformations separate malicious semantics from superficial code
form. A representation associated with malicious logic should change
substantially between $x$ and $x_{\mathrm{cf}}$, while remaining stable
between $x$ and $x_T$. This example explains why critical layer localization
jointly evaluates malicious logic sensitivity and surface perturbation
stability.


\begin{table}[!t]
\centering
\caption{
\textbf{Illustration of malicious logic removal and surface perturbation.}
The original input contains an unsafe transfer restriction. Malicious logic
removal eliminates this restriction and changes the target label from
malicious to benign. The behaviour-preserving surface perturbation renames
identifiers while retaining the unsafe logic and malicious label. The excerpts
are abridged from variants that pass Solidity 0.5.17 compilation.
}
\vspace{-0.4cm}
\label{tab:transformation_case}

\footnotesize
\setlength{\tabcolsep}{3pt}
\renewcommand{\arraystretch}{1.0}

\begin{tabularx}{\columnwidth}{@{}X@{}}
\toprule

\rowcolor{gray!15}
\textbf{Original malicious smart contract} \hfill \textit{Original input: $x$} \\
\midrule

\begin{minipage}[t]{\linewidth}
\scriptsize\ttfamily
\soltype{address} \solkw{private} blockedTarget;\par
\solkw{function} setBlockedTarget(\soltype{address} target)\par
\hspace*{1em}\solkw{external} onlyOwner \{\par
\hspace*{1em}blockedTarget = target;\par
\}\par
\solunsafe{// unsafe: owner selects the blocked target}\par
\solkw{require}(to != blockedTarget, \solstr{"please wait"});\par

\normalfont\footnotesize\vspace{1pt}
\textbf{Unsafe behaviour:}
The owner selects an address that is prevented from receiving token transfers.
\par\noindent
\textcolor{red!70!black}{
\textbf{(unsafe logic present $\rightarrow$ malicious/honeypot, label = 1)}
}
\end{minipage}
\\

\addlinespace[1pt]
\midrule

\rowcolor{gray!15}
\textbf{Malicious logic removal} \hfill \textit{Counterfactual input: $x_{\mathrm{cf}}$} \\
\midrule

\begin{minipage}[t]{\linewidth}
\scriptsize\ttfamily
\solkw{function} setBlockedTarget(\soltype{address} target)\par
\hspace*{1em}\solkw{external} onlyOwner \{\par
\hspace*{1em}target = target;\par
\}\par
\solcomment{// owner-controlled state update removed}\par
\solkw{require}(\solkw{true}, \solstr{"please wait"});\par
\solcomment{// transfer restriction removed}\par

\normalfont\footnotesize\vspace{1pt}
\textbf{Semantic change:}
The owner-controlled state update and the blocked-address restriction are
removed.
\par\noindent
\textcolor{green!45!black}{
\textbf{(unsafe logic removed $\rightarrow$ benign, label = 0)}
}
\end{minipage}
\\

\addlinespace[1pt]
\midrule

\rowcolor{gray!15}
\textbf{Behaviour-preserving surface perturbation} \hfill \textit{Perturbed input: $x_T=T(x)$} \\
\midrule

\begin{minipage}[t]{\linewidth}
\scriptsize\ttfamily
\soltype{address} \solkw{private} restrictedAddress;\par
\solkw{function} setBlockedTarget(\soltype{address} candidate)\par
\hspace*{1em}\solkw{external} onlyOwner \{\par
\hspace*{1em}restrictedAddress = candidate;\par
\}\par
\solcomment{// identifiers renamed only}\par
\solkw{require}(to != restrictedAddress, \solstr{"please wait"});\par
\solunsafe{// unsafe: transfer restriction unchanged}\par

\normalfont\footnotesize\vspace{1pt}
\textbf{Surface change:}
The identifiers are renamed while the owner-controlled transfer restriction
remains active.
\par\noindent
\textcolor{red!70!black}{
\textbf{(unsafe logic retained $\rightarrow$ malicious/honeypot, label = 1)}
}
\end{minipage}
\\

\bottomrule
\end{tabularx}
\vspace{-0.4cm}
\end{table}

\section{More Discussions}
\label{app:dis}

\noindent $\triangleright$ \textbf{\textit{Do MCDUs capture latent security knowledge or merely classification correlations?
}} \\
High classification performance alone does not establish that the identified
units encode malicious logic, since a detector may rely on superficial cues
such as identifiers, comments, or code templates. ContractLens therefore
introduces semantic controls at both the layer and neuron levels. We require
the selected representations to respond to malicious-logic removal while
remaining stable under behaviour-preserving surface perturbations. In addition, masking the identified MCDUs causes substantially larger performance degradation than masking randomly selected or activation-matched neurons. Together, these results suggest that MCDUs capture functionally relevant signals associated with malicious contract semantics rather than arbitrary classification
correlations.

\noindent $\triangleright$ \textbf{\textit{Why are multiple signals required for MCDUs identification?
}} \\
Each signal captures a different property of a candidate neuron. Activation
contrast measures whether its response differs between malicious and benign
contracts, while the probe weight measures its contribution to the learned
decision boundary. Malicious-logic specificity further distinguishes neurons
that respond to changes in malicious semantics from those primarily sensitive to superficial code variations. Combining these complementary signals therefore favours neurons that are simultaneously discriminative, decision-relevant, and semantically specific, rather than neurons that score highly under only one criterion.

\noindent $\triangleright$ \textbf{\textit{Can the identified security knowledge generalize beyond malicious behaviours observed during training?
}} \\
Our experiments indicate that the identified representations exhibit meaningful, although not universal, cross-behaviour generalization. In the leave-one-intent-out evaluation, all localization and training stages exclude the held-out malicious intent, which is used only for testing. ContractLens still achieves higher average performance than the evaluated baselines. We further evaluate contracts deployed after the release of the backbone model and observe strong detection performance on several malicious mechanisms. These results suggest that the localized representations capture reusable characteristics of malicious contract logic rather than only memorizing individual attack categories. Nevertheless, ContractLens should not be interpreted as a universal detector for arbitrary future attacks, and its coverage remains dependent on the security knowledge represented by the underlying pretrained model.

\end{document}